\documentclass[aip,pop,preprint,groupeaddress]{revtex4-1}
\usepackage{graphicx}
\usepackage{dcolumn,amsmath,amssymb}
\usepackage{bm}
\usepackage{multirow}
\begin{document}
\title{\normalsize\textbf{Interplay of transitions between oscillations with emergence of fireballs and quantification of phase coherence, scaling index
in a magnetized glow discharge plasma of toroidal assembly}}
\author{Debajyoti Saha}
\email{debajyoti.saha@saha.ac.in}
\affiliation{Plasma Physics Division, Saha Institute of Nuclear Physics, HBNI, Bidhannagar, Kolkata - 700064, India.}
\author{Sabuj Ghosh}
\affiliation{Plasma Physics Division, Saha Institute of Nuclear Physics, HBNI, Bidhannagar, Kolkata - 700064, India.}
\author{Pankaj Kumar Shaw}
\affiliation{Plasma Physics Division, Saha Institute of Nuclear Physics, HBNI, Bidhannagar, Kolkata - 700064, India.}
\author{M.S. Janaki}
\affiliation{Plasma Physics Division, Saha Institute of Nuclear Physics, HBNI, Bidhannagar, Kolkata - 700064, India.}
\author{A.N. Sekar Iyengar}
\affiliation{Plasma Physics Division, Saha Institute of Nuclear Physics, HBNI, Bidhannagar, Kolkata - 700064, India.}

\begin{abstract}
Interplay of transition of floating potential fluctuations in a glow discharge
plasma in the toroidal vacuum vessel of SINP tokamak has been observed.
With variation in the strength of the vertical and toroidal magnetic fields, regular and inverted relaxation oscillations as well as sinusoidal oscillations are observed with the slow and fast time scale of the relaxation oscillations reversing their nature at a high value of vertical magnetic field strength.
However for small value of toroidal magnetic field the transitions follow relaxation $\rightarrow$  chaotic oscillations
with the chaotic nature prevailing at  higher values of toroidal magnetic field. Evolution of associated anode fireball dynamics under the action of increasing vertical, toroidal  as well as increasing vertical field at a fixed toroidal field (mixed field) of different strength has been studied.
Estimation of phase coherence index for each case has been carried out to examine the evidence of finite nonlinear interaction.
A comprehensive study of the dynamics of the fireball is found to be associated with the values of phase coherence index. The index is found to take maximum values for the case of toroidal, mixed field when there is an existence of power/energy concentration in a large region of frequency band. A detailed study of the scaling region using detrended fluctuation analysis (DFA) by estimating the scaling exponent has been carried out for increasing values of discharge voltage, vertical, toroidal as well as the mixed field (toroidal plus vertical). A persistence long range behaviour associated with the nature of the anode glow has been investigated in case of higher values of toroidal, mixed field whereas increasing DV, vertical magnetic field leads to a perfectly correlated dynamics with values of scaling exponent greater than unity.

\end{abstract}
\maketitle

\section{Introduction}

Fireballs \cite{fire1,vramori_homoclinic} are studied in a glow discharge device consisting of cathode biased negatively
with  respect to grounded chamber well. When the positively biased electrode is immersed
a glow around the electrode is observed. Anode fireballs are discharge phenomena near
the positively biased electrode. These are highly nonlinear phenomena involving the physics of sheaths , double layer \cite{sujit}, ionization beam
and possibly external circuit interaction. Fundamental questions remains with respect to the peculiar shape of the fireballs, the physics of relaxation oscillation \cite{firerel} including waves and instabilities created by the non-Maxwellian distribution function both in magnetized and unmagnetized plasma.
A steady state double layer requires a momentum balance \cite{fire1, firerel} or flux ratio $\frac{J_e}{J_i}= \sqrt{\frac{m_i}{m_e}}$ .  The lack of the pressure or momentum balance would not lead to a stationary configuration. In many situations fireball grows but does not reach equilibrium due to unequal ion production and losses resulting in a repetitively pulsating fireball \cite{fire2}. There are recent experiments conducted to address the observation of multiple double layers associated with the anode glows in toroidal geometry \cite{paul}. Much of the attention has been on the formation of double layers \cite{double}and relaxation oscillation \cite{arun_homoclinic} of unstable fireballs which have been analyzed in the framework of chaos theory \cite{Ionita}. In addition to the investigation of the dynamics of the fireball \cite{fire3}, here we study the association of the glow with some quantitative aspects like phase coherence index and scaling index. We attempt to find the correlations among phases within the context of surrogate data \cite{sch} in a magnetized glow discharge plasma in a toroidal assembly \cite{sabuj_toroidal} and relate the phase coherence indices with the different shape of the glow at different parametric regimes.

Plasmas are intrinsically nonlinear whose effects manifest in the form of various exotic structures such as double layers, solitons, vortices,
different types of waves, instabilities and turbulence \cite{lib}. Glow discharge plasma being rich in high energy, electrons and ions are capable of exhibiting many such nonlinear phenomenon\cite{vramori_MMO, mypaper,jaman_chaos}. Much effort has been endeavoured to study the intricacies involving the topics like finite nonlinear interactions and its associated phase coherence index \cite{koga, hada} in magnetohydrodynamic turbulence in solar wind. Here in our study we address for the first time the evidence of finite nonlinear interaction for increasing discharge voltage (DV), vertical, toroidal magnetic field ($B_V,B_T$) and the mixed field ($B_{V+T}$) in the toroidal vacuum vessel of SINP tokamak. In order to investigate the nonlinear wave wave interactions we need to decompose the signal. From this point of view phase information obtained from the Fourier transform is still convenient for our analysis because it permits wave number/frequency decomposition albeit we have to implicitly assume weak nonlinearity. This is the first time that the distribution of wave phases for two different kinds of surrogate data namely phase randomised (PRS), phase constant (PCS) along with the original one has been executed.

In the last one decade detrended fluctuation analysis (DFA) \cite{peng,peng1} has emerged as an important technique to study scaling and long range
temporal correlation in a nonstationary time series \cite{hamilton} which has been extensively studied in literature. DFA is based on the idea that  if the time series contains nonstationarities then the variance of the fluctuations can be studied by successively detrending using linear quadratic, cubic higher order polynomial in a piecewise manner. Most real time series exhibit persistence i.e subsequent element of the time series are correlated \cite{jamanda_SOC,jamanda,pankaj}. Fluctuations in the indices of major stock market are long range correlated  physically implying that the correlations in the fluctuations do not decay fast and the system possesses long memory. This property is also true for EEG series, temperature anomalies worldwide.
This arises from very specific kinds of correlation existing among the fluctuations after the trends are removed.
The study of  self similarity and scaling in physics and socio economic sciences in the last several years has brought in new insights and new ideas for modeling them. For instance one of the important empirical results of the market dynamics is that the probability distribution of price returns r in a typical market displays a power law \cite{DFA1} i.e $P(r)\sim r^{\alpha}$ where $\alpha$ =3. Similar power laws appear for the cumulative frequency distribution of earthquake magnitudes \cite{DFA2}.

The paper is organized as follows. In section II, we present a brief schematic of the experimental setup, followed by the results of the analysis of floating potential fluctuation (FPF) and power spectral analysis in section III.  A comprehensive analysis of the evolution of fireball dynamics with the parameters like DV, $B_V, B_T$, $B_{V+T}$ followed by the results of the analysis of path length, phase coherence index has been illustrated respectively in section IV and V. Section VI demonstrate the results of the analysis for DFA method to estimate the scaling exponent and its association with the fireball nature. Conclusions are presented in section VII.

\section{Experimental Setup}

\begin{figure}
\centering
\includegraphics[width=9cm,height=6cm]{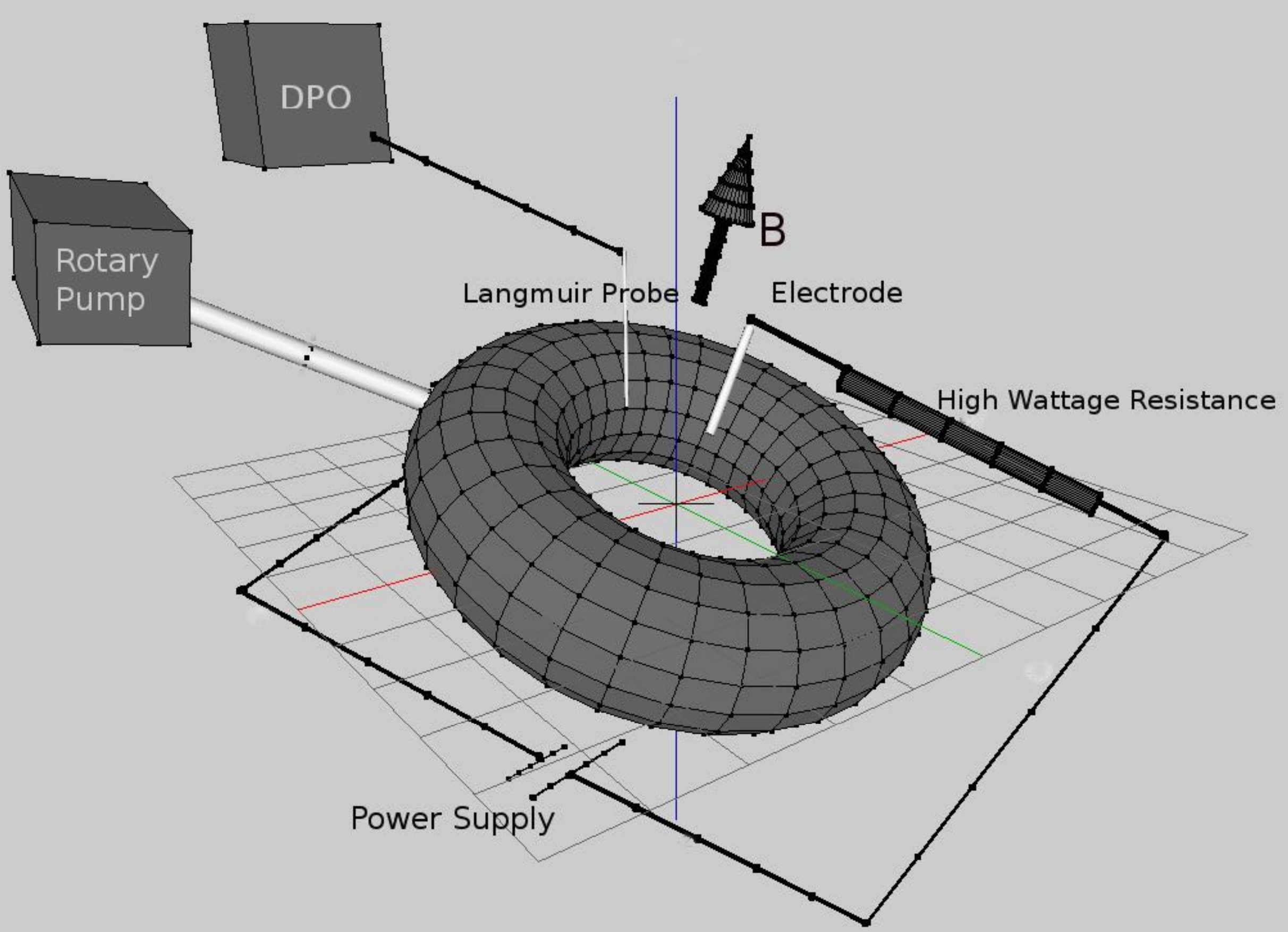}
\caption{Schematic diagram of the experimental setup}
\label{exp}
\end{figure}

The toroidal vessel of SINP tokamak with major radius 30cm and minor radius 7.5 cm has been used as the
discharge chamber. The grounded vessel was evacuated to a base pressure of 0.01 mbar using a rotary
oil pump. The vessel was filled with hydrogen upto a pressure of 0.45mbar and a discharge voltage (DV)upto 0.60kV to sustain
the discharge plasma. A Langmuir probe was inserted to measure the floating potential fluctuations in the system. An electrode of
length 3 cm and mean radius 0.7cm was used as an anode while the vacuum vessel was grounded as shown in the Fig. \ref{exp}. Four coils, two above the torus and two below the torus having 24 turns each were used to create the vertical magnetic field in the vessel whereas 16 coils having 12turns/coil were utilised in generating toroidal magnetic field with 1 Ampere current in the coil producing 1.28Gauss of $B_T$. The direction of B in the Fig. \ref{exp} is same as that of vertical magnetic field $B_V$.  The experiments were conducted under the action of increasing DV, vertical, toroidal magnetic field, and increasing vertical magnetic field at a fixed toroidal magnetic field of 8.5Gauss.

\section{Floating potential fluctuation, power spectral analysis}

\subsection{Increasing DV, vertical and toridal magnetic field ($B_V, B_T$)}

With increase in DV the nature of the transitions is observed to be of homoclinic type upto 0.42kV. Beyond this value, the FPF's again become
relaxation type which is seen to change to regular periodic behaviour at 0.46kV. This nature persisted upto 0.50kV after
which some irregularities appear in the regular sinusoidal behaviour. So the transitions undergo a homoclinic behaviour and subsequently modify to
regular periodic oscillation followed by the irregular type at high values of DV.

\begin{figure}
\includegraphics[width=8.6cm, height=5cm]{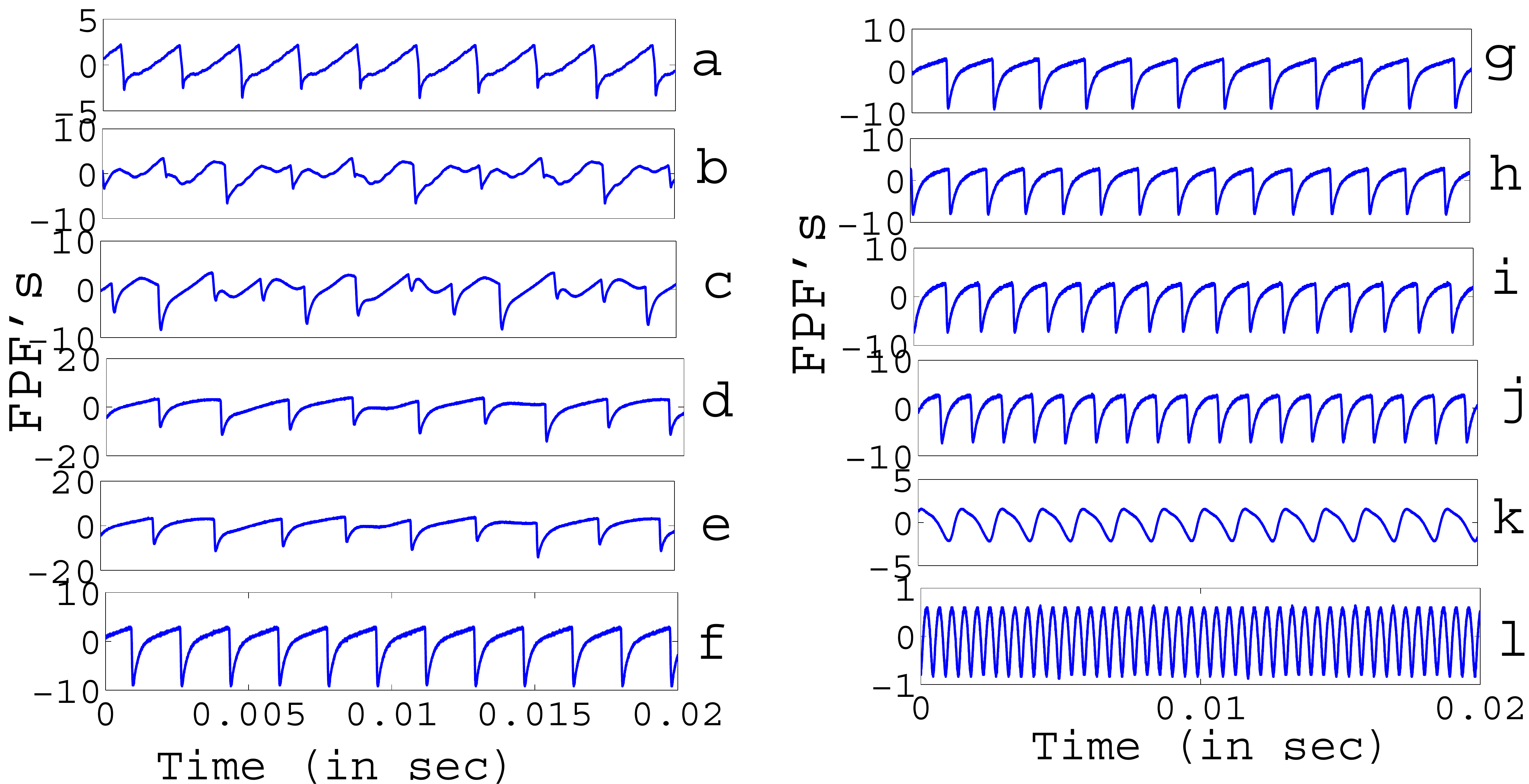}
\includegraphics[width=8.6cm, height=5cm]{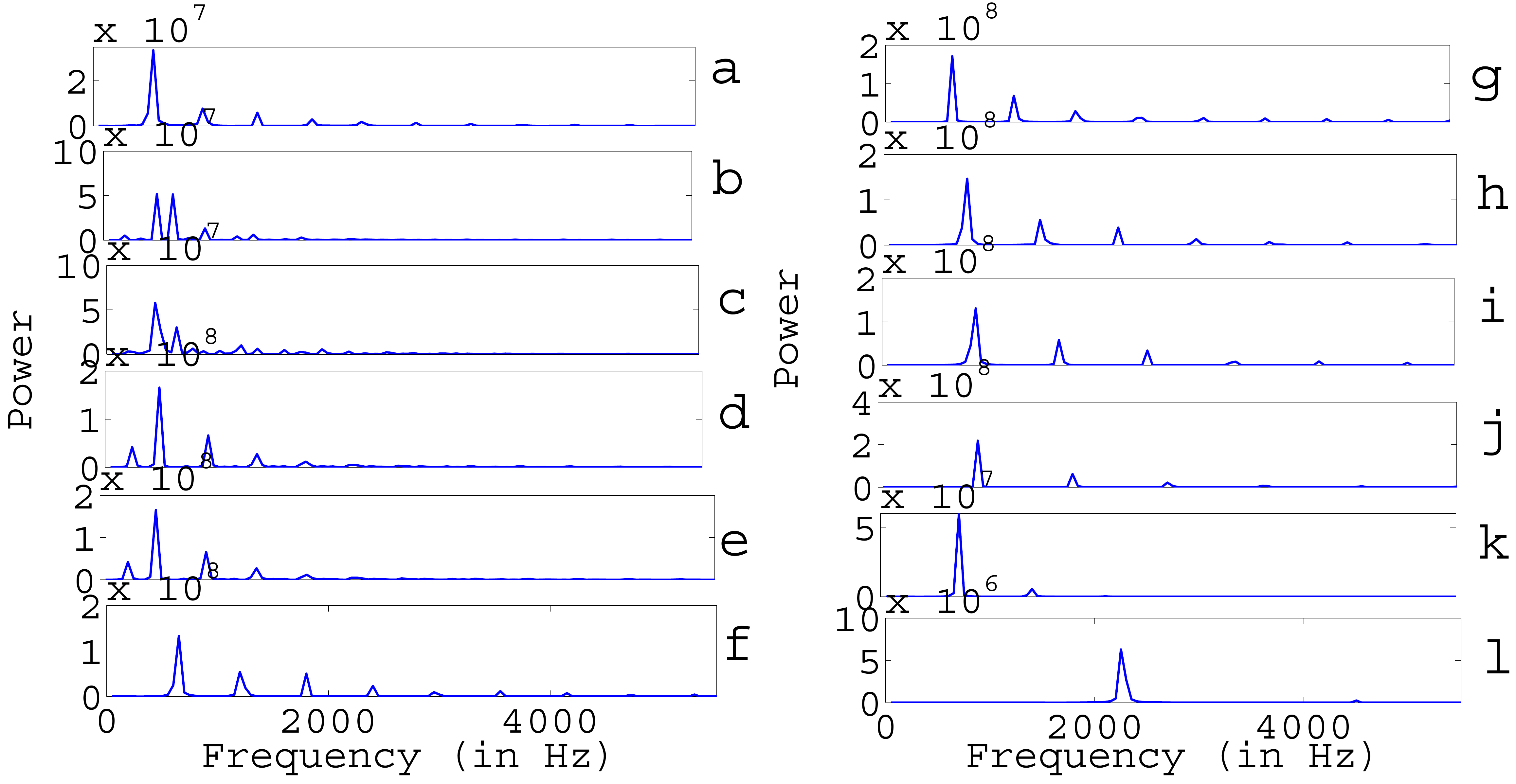}
\caption{Plot of FPF's and power spectrum with the variation in $B_V$ a)0G b) 3.36G c) 3.65G d) 4.21G e) 6.17G f) 7.29G  g) 7.57G  h) 8.70G i) 9.54G j) 10.11G k) 11.23G  l) 11.79G}
\label{rawver}
\end{figure}

Now keeping the DV at a value of 0.44kV we apply vertical magnetic field upto $B_{V}$=11.8 Gauss (G).
The plasma floating potential fluctuations were found to show oscillatory features of different type depending on the intensity of the applied $B_{V}$ (Figure \ref{rawver}). When the value of $B_V$ is increased, the nature of the relaxation oscillations persisted but the rising time scale of the relaxation oscillation get steeper at the onset of $B_V$=7.29G and sustain upto 10.105G as observed from figure \ref{rawver}. The change of nature of the rising and decay time scale of the relaxation oscillation is worth observing at $B_V$=11.23G (Fig. \ref{rawver}k).

\begin{figure}
\includegraphics[width=8.6cm, height=5cm]{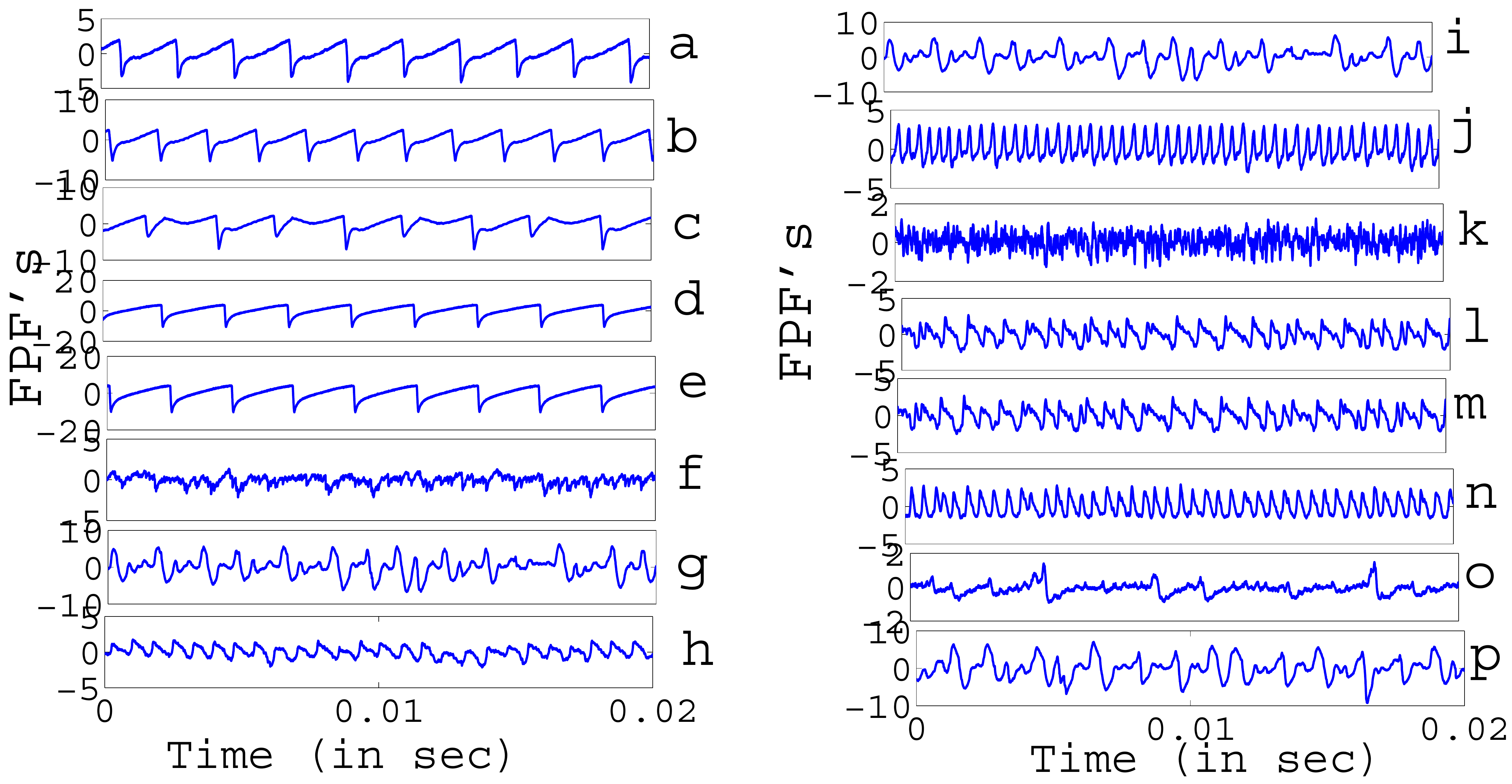}
\includegraphics[width=8.6cm, height=5cm]{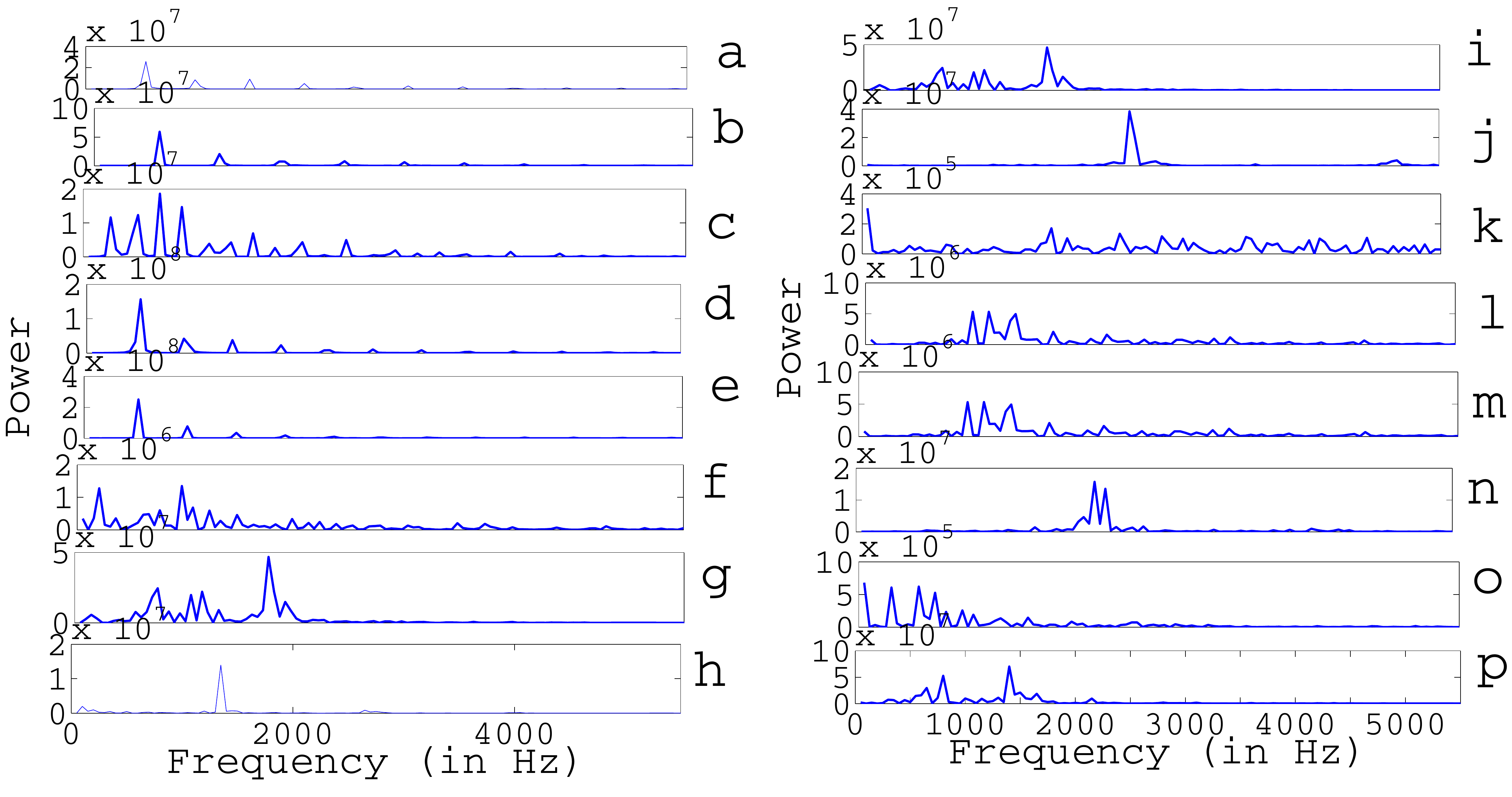}
\caption{Plot of FPF's and power spectrum with the variation in $B_T$ in the left and right panel respectively: a)0G b) 1.28G c) 2.30G d) 3.07G e) 5.63G f) 7.098G  g) 8.06G h) 8.76G i) 8.98G j) 9.6G  k) 10.24G l) 11.2G m) 11.52G n) 11.84G  o) 12.8G p) 14.4G}
\label{rawtorr}
\end{figure}

\begin{figure}
\includegraphics[width=8.6cm, height=5cm]{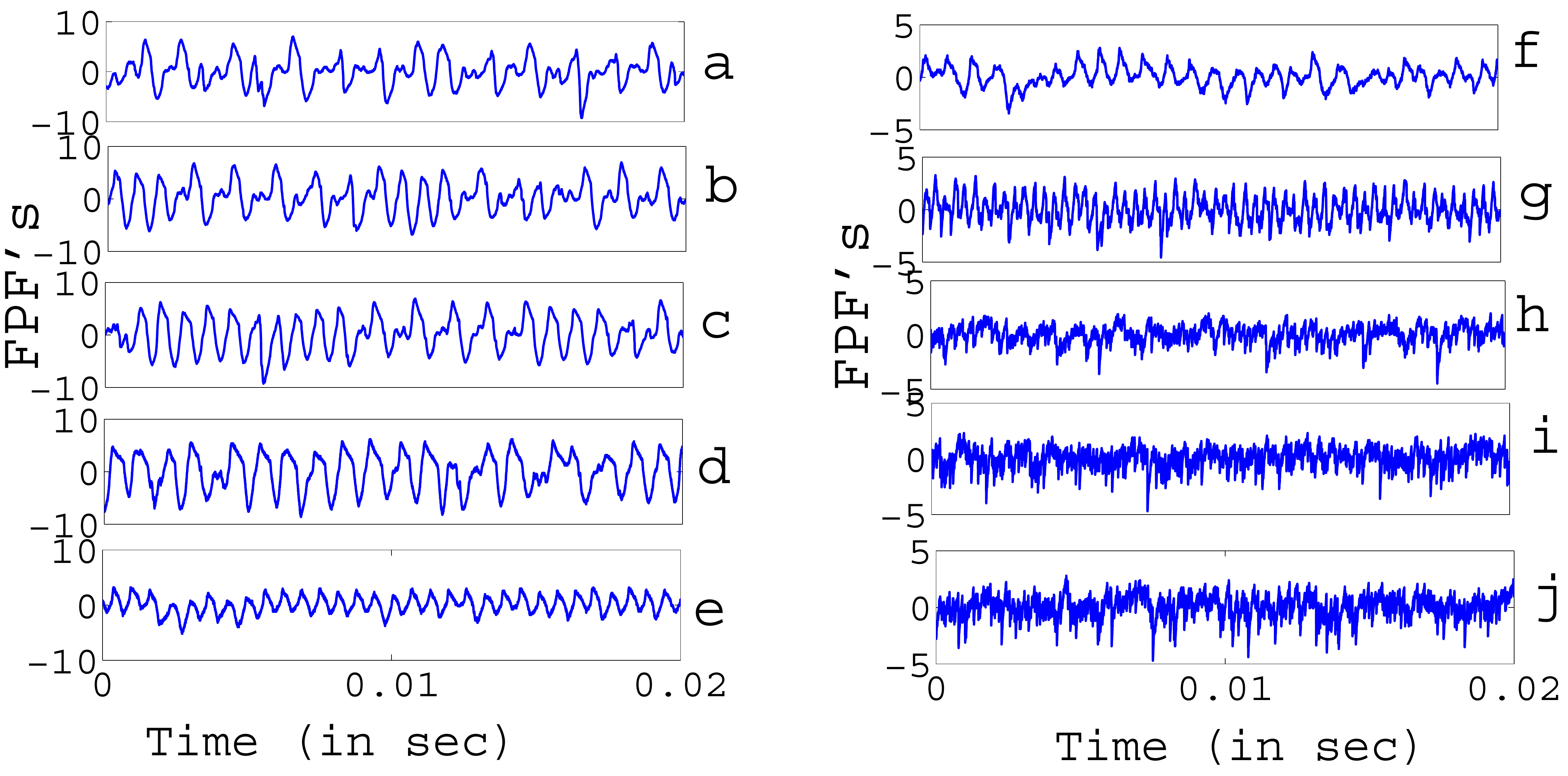}
\includegraphics[width=8.6cm, height=5cm]{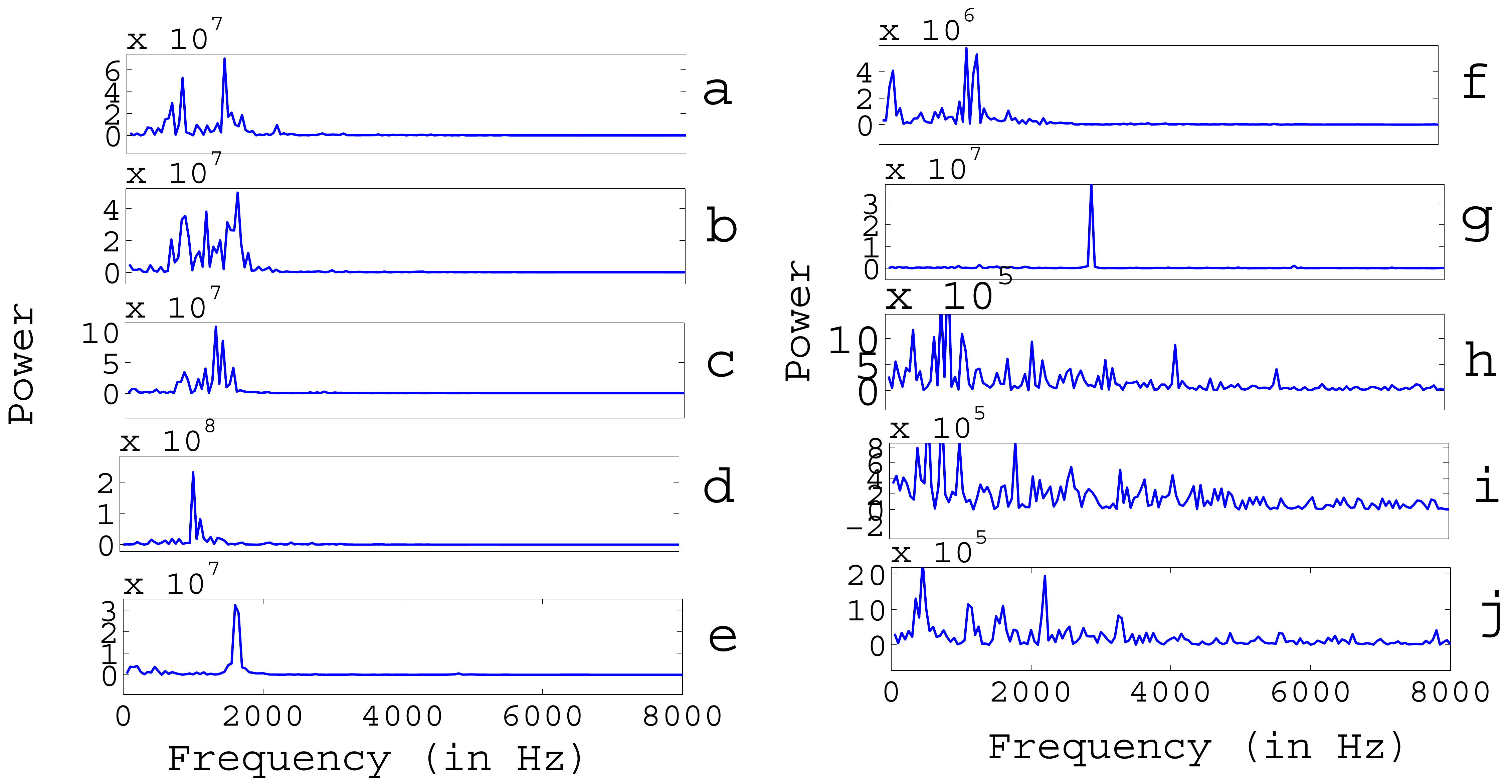}
\caption{Sequential change in the FPF's and frequency spectrum for different $B_T$ for fixed $B_T$= G in the left and right panel respectively: a) 0G
b) 0.84G c) 1.54G d) 2.47G e) 4.09G f) 5.61G  g) 7.02G h) 8.98G i) 10.67G j) 12.63G}
\label{rawmix}
\end{figure}

The transitions with the application of toroidal magnetic field ($B_T$) exhibit a drastic change. Similar to the $B_V$ here also the emergence of relaxation oscillation with slow and fast time scale takes place upto $B_T$=5.63G. After that the system undergoes a chaotic behaviour with a sudden emergence of broadband characteristics which is prominent if we apply vertical field keeping the toroidal field at a fixed value portrayed in Fig.\ref{rawmix} h-j. The chaotic nature of the FPF's for increasing $B_T$ has been depicted in Fig. \ref{rawtorr} f-p. At low values of $B_T$ the amplitudes of the chaotic oscillation are seen to be very low which get increased with the increase in $B_T$(Fig \ref{rawtorr}). When $B_T$ reaches 11.2G, we can observe the emergence of some sort of relaxation oscillation having the chaotic nature persisting upto 11.84G as seen in the right panel of Fig. \ref{rawtorr}. Further change in the nature of the relaxation oscillations takes place at a high value of $B_T$ $\sim$ 12.8G. Finally we try to observe the effects of applying $B_V$ keeping the toroidal field fixed at a value of 8.56G. All the oscillations are observed to reveal chaotic nature with prominent broadband characteristics for higher value of mixed field as depicted in Fig. \ref{rawmix}.

\begin{figure}
\includegraphics[width=14cm, height=9cm]{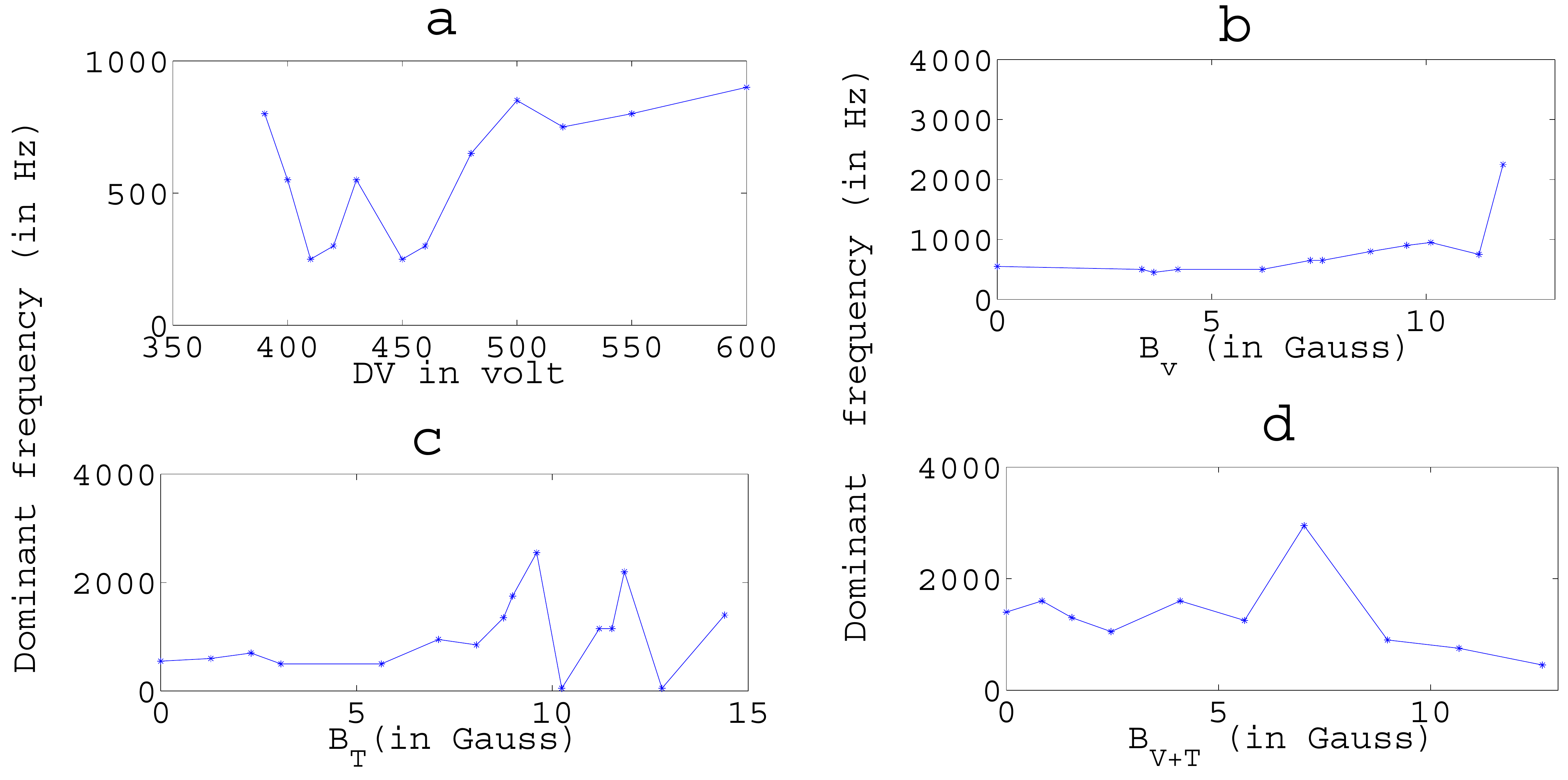}
\caption{Dominant frequency with the variation in a) Discharge voltage(DV) b) Vertical magnetic field ($B_v$) c) Toroidal magnetic field ($B_T$) d) Vertical field at fixed $B_T$ of 8.45G ($B_{V+T}$)}
\label{freqdom}
\end{figure}

The frequency carrying maximum power (dominant frequencies) with the variation in DV, $B_V, B_T$ have been plotted in Fig. \ref{freqdom}. The dominant frequencies with DV are seen to lie from 150 $\sim$ 900 Hz. A minimum frequency of 250 Hz has been obtained at DV $\sim$ 450V followed by the increase in dominant frequencies upto 500V where maximum dominant frequency of 850 Hz have been achieved. The application of vertical field gradually enhances the frequencies after 6G and a sharp increase in the value of main frequency is noted at $B_V$ $\sim$ 11.79 Gauss. Toroidal magnetic field has impact in changing the frequency values to a large extent in a quite random way. A range of dominant frequencies(850, 1750, 2200, 2550, 1150) Hz have been noted with a prominent maxima at 2550 Hz at $B_T$=9.6G along with the presence of a local maxima of 2200Hz at 11.84G. The values of the main frequencies with the application of increasing vertical field on a fixed $B_T$(8.56G) remain from minimum of 450Hz to maximum of 2950 Hz with the maximum frequency being achieved at around $B_V=7.01$ Gauss. Further increase in the values of ($B_{V+T}$) the frequency with maximum power shows a gradual decreasing trend.

\begin{figure}
\includegraphics[width=8.6cm, height=5cm]{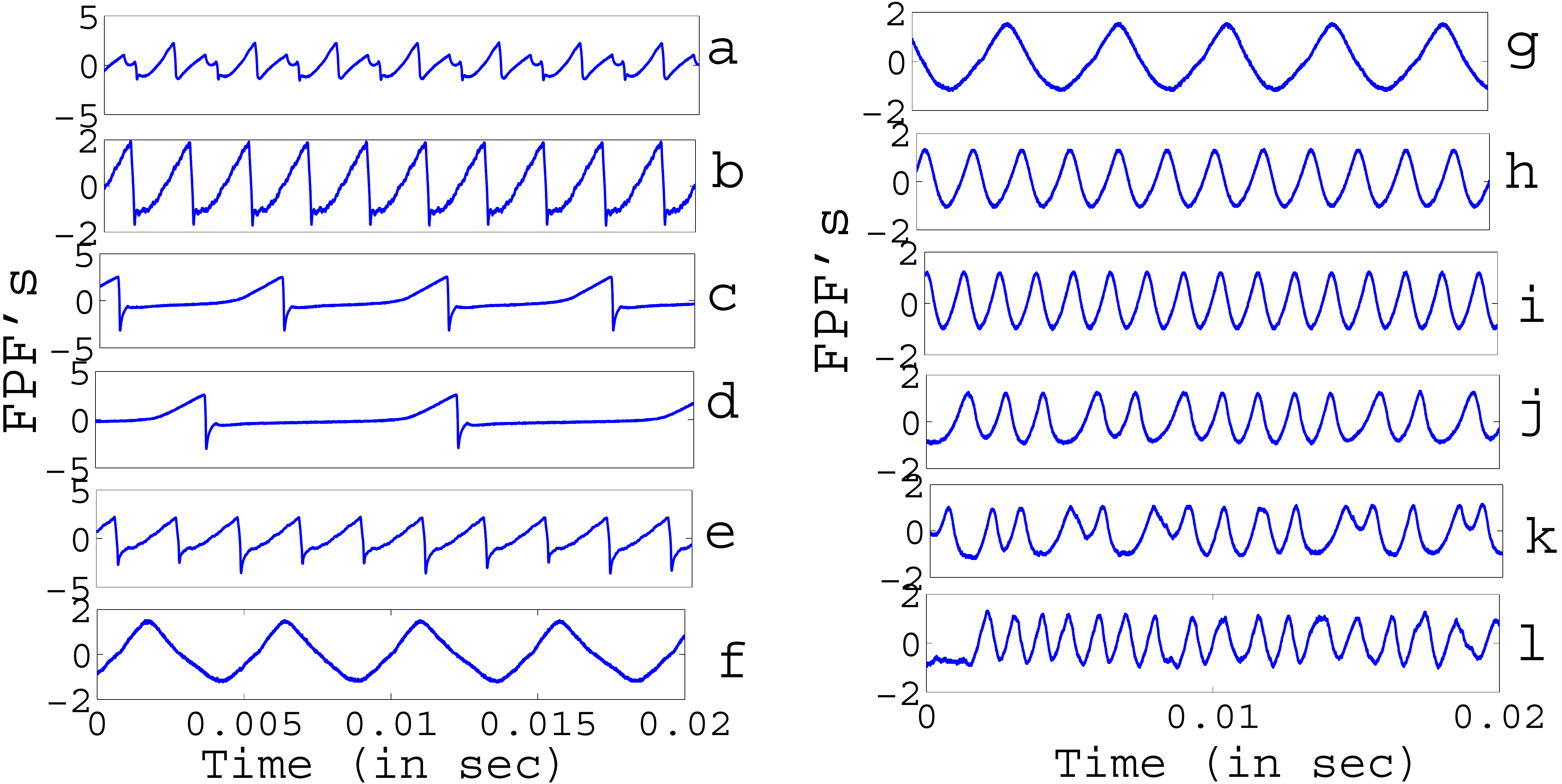}
\includegraphics[width=8.6cm, height=5cm]{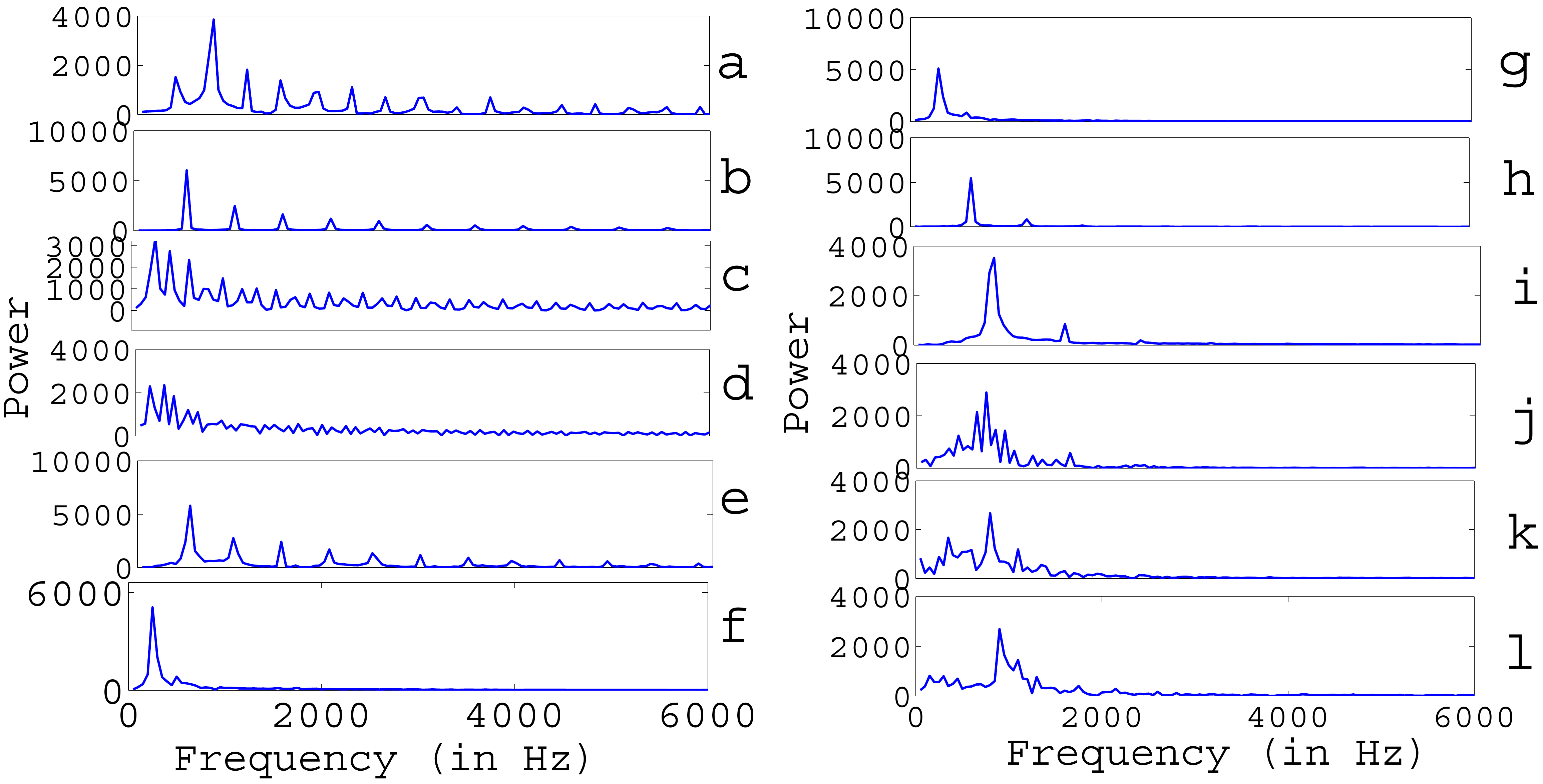}
\caption{Sequential change in the raw signals and frequency spectrum for different DV (in Volts) in the left and right panel respectively:
a) 390V b) 400V c) 410V d) 420V e) 430V f) 450V g) 460V h) 480V i) 500V j) 520V k) 550V l) 600V}
\label{rawdv}
\end{figure}

\section{Evolution of Fireballs with DV, \textbf{$B_V, B_T, B_{V+T}$}}

\begin{figure}
\centering
\includegraphics[width=11cm]{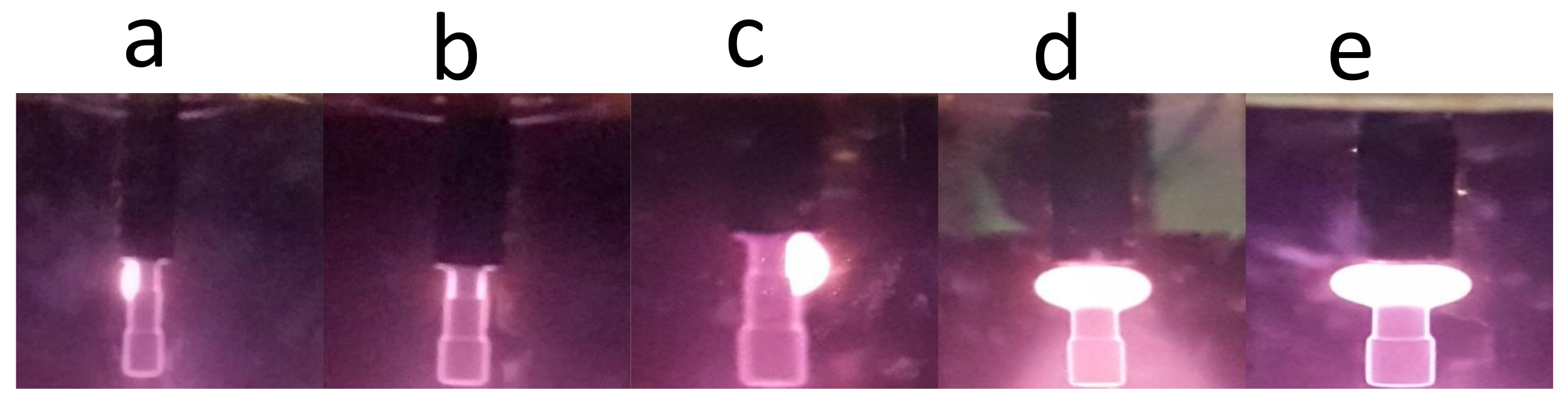}
\caption{Photograph of the anode glow with variation in DV a) 390V b) 400V c) 420-440V d) 460V e) 520V}
\label{glow}
\end{figure}

The sheath formation and its dynamics around the electrode is an important phenomenon
in a glow discharge system since it contributes to the various frequency components. The glow indicates the energetic electrons impacting with the neutrals.
The electrode collects electrons from the plasma as a result of which the anode is encapsulated, partially or fully,
by ions and quasi-neutrality of plasma leads to the formation of double layers.
To start with, when the plasma was formed a small fireball was found to be attached to the anode at a particular position (here left side) as depicted in
Fig \ref{glow}. With the emergence of relaxation oscillation the position of the glow was found to rotate along clockwise direction aligning itself to the back side of the electrode for DV=0.40kV and then shift to the right side at DV=0.42kV persisting upto DV=0.44kV in presence of homoclinic transition.
Then the position of the fireball adjust itself to encircle the electrode for DV=0.46-0.52kV when FPF's display regular periodic oscillations.

\begin{figure}
\centering
\includegraphics[width=10cm]{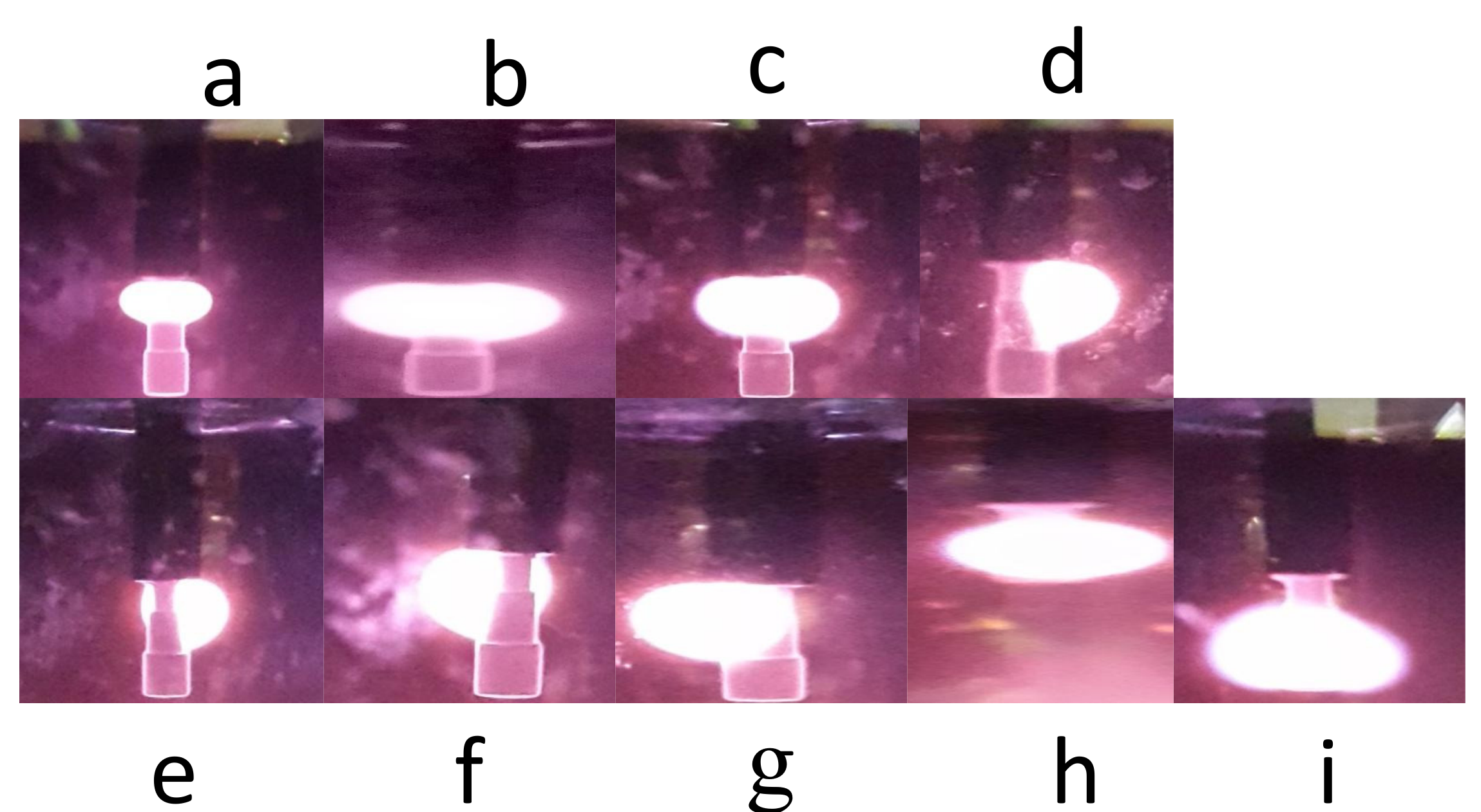}
\caption{Sequential change in the anode glow  for increasing vertical magnetic field $B_V$ a) 3.65G b) 4.21G c) 6.18G  d) 7.29G e) 8.70G f) 9.54G g) 10.10G h) 11.23G i) 11.79G}
\label{glow1}
\end{figure}

The nature of the glow for zero magnetic field has already depicted in Fig. \ref{glow}d for DV=0.44kV. For initial values of ($B_V$) upto 6.18G the location of the fireball was on both sides of the electrode. The shape of the glow in Fig. \ref{glow1}b is little elongated corresponding to the time series at $B_V$=4.21G showing relaxation oscillation with increasing rise time in comparison to that observed at the beginning. When the rise time scale of the relaxation oscillation get steeper, fireball aligns itself only to the right side of the electrode (Fig \ref{glow1} d) i.e the fireball undergoes rapid relaxation oscillation. The anticlockwise rotation of the position of the fireball (Fig. \ref{glow1} e,f,g) is observed when the slope of the rise time scale is again slightly changed. Most interestingly, on complete reversal of the nature of the time scale of the relaxation oscillation ($B_V$=11.23G) (Fig. \ref{rawver}k) a comparatively broader shaped fireball near the electrode was observed in Fig \ref{glow1}h which is seen to persist its shape only with slight increase in intensity when regular periodic oscillation at $B_V$=11.79G appears in Fig. \ref{glow1} i.

\begin{figure}
\centering
\includegraphics[width=10cm]{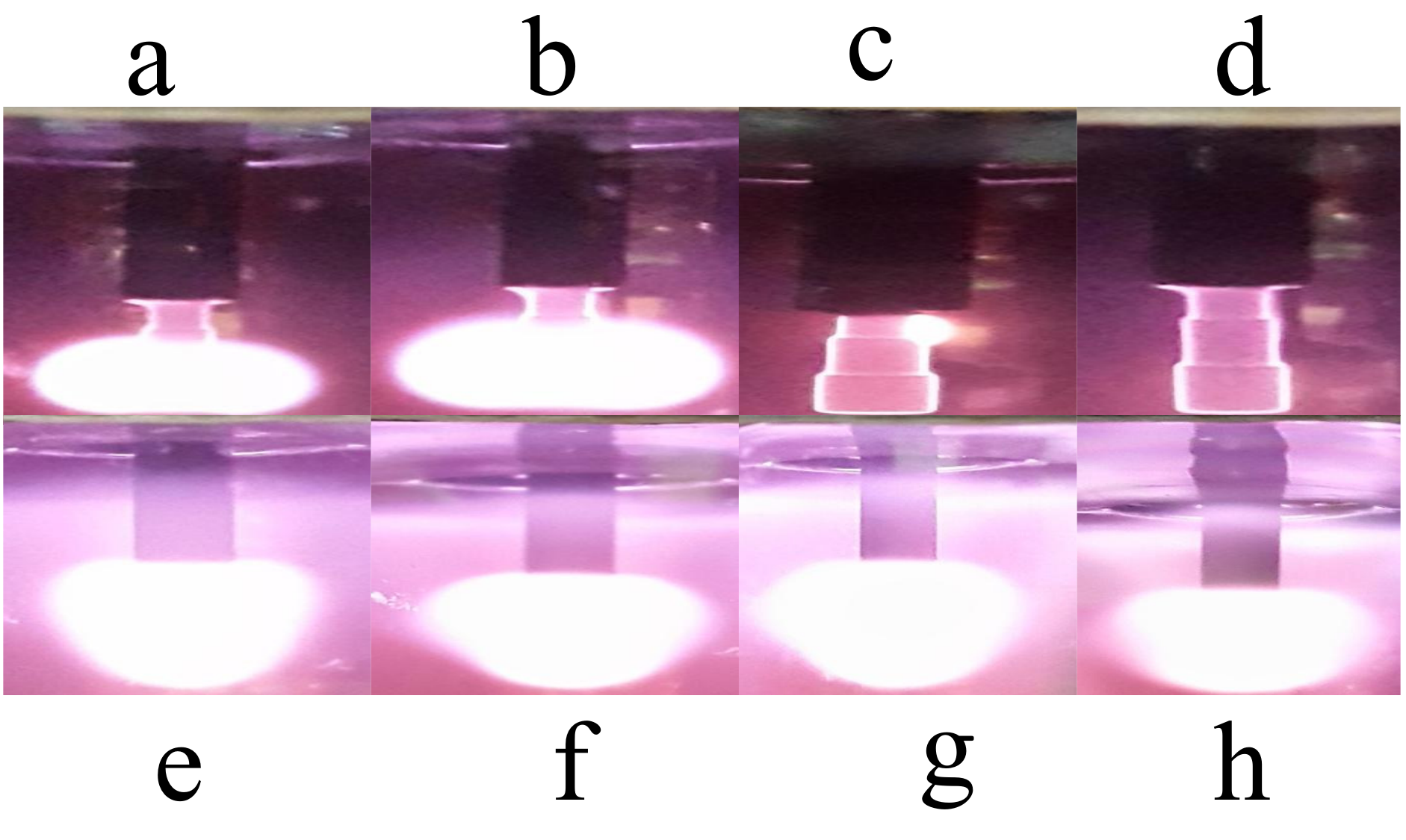}
\caption{Sequential change in the glow for increasing toroidal magnetic field $B_T$ a) 1.28-2.30G b) 3.07-5.63G c) 7.09G d) 8.06-8.98G e) 9.6G f) 10.24G g) 11.2-11.84G h) 11.84-14.4G }
\label{glow2}
\end{figure}

\begin{figure}
\centering
\includegraphics[width=10cm]{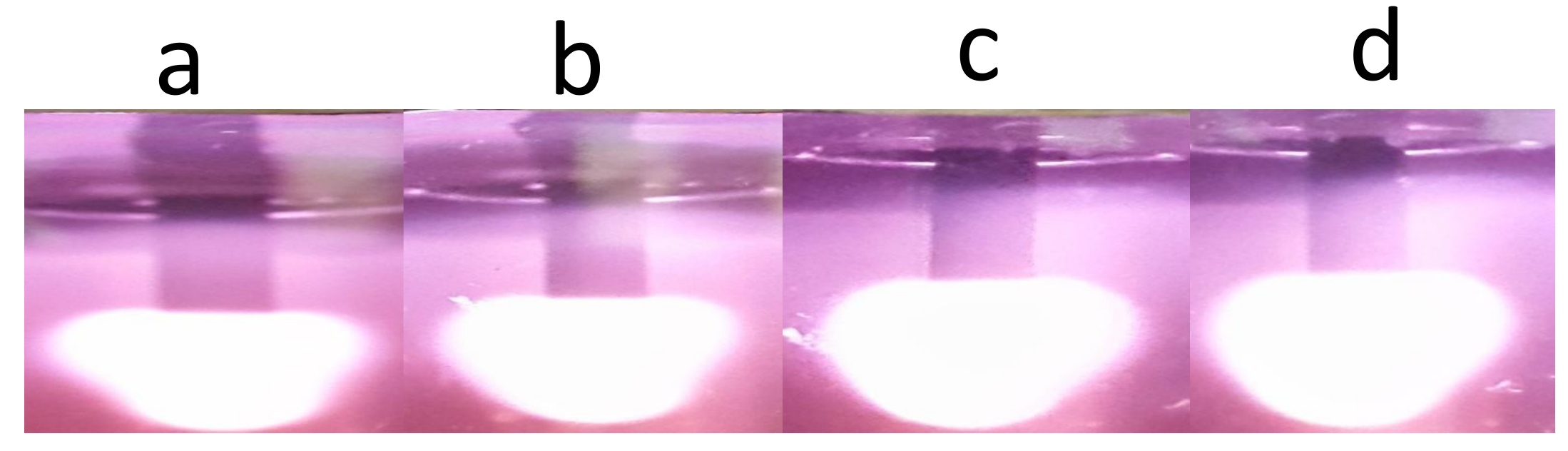}
\caption{Photographs of the anode glow for increasing $B_{V+T}$ a)0.84G-2.47G  b) 4.09-7.02G c) 8.98-10.67G d) 12.63G}
\label{glow3}
\end{figure}

A some sort of elliptical shaped nature of the fireballs was persistent for the relaxation oscillation generated in presence of low toroidal field upto 5.63G. The size of glow in Fig \ref{glow2}b is observed to occupy larger space than that in Fig \ref{glow2}a. With the emergence of chaotic oscillations at intermediate values of toroidal field (upto $B_T$=8.96G), a self rotation of the glow takes place and finally the fireball is seen to take position on the back side of the electrode as seen in Fig \ref{glow2}d. The appearance of relaxation oscillation bearing the chaotic nature lead to the generation of a completely different shaped glow compared to the previous ones( Fig. \ref{glow2}e). This particular fireball shape is reminiscent of the inverted pear type. On further increase in $B_T$ the inverted pear shaped nature prevails along with the increase in intensity of the glow for FPF's undergoing chaotic nature (Fig. \ref{glow2}f-h). However while the system is subjected to increasing vertical field upto $B_V$=2.47G on a fixed toroidal field i.e mixed field $(B_{V+T})$ , the shape of the glow appears to become little distorted followed by more localisation in intensity (Fig. \ref{glow3}b). Gradual increase in $B_{V+T}$ leads to the slight increase in plasma density along with the fireball size as checked by visual inspection from Fig. \ref{glow2} c-d.

\section{Method of analysis: Path length, Phase coherence index}

\begin{figure}
\centering
\includegraphics[width=4.5cm, height=3cm]{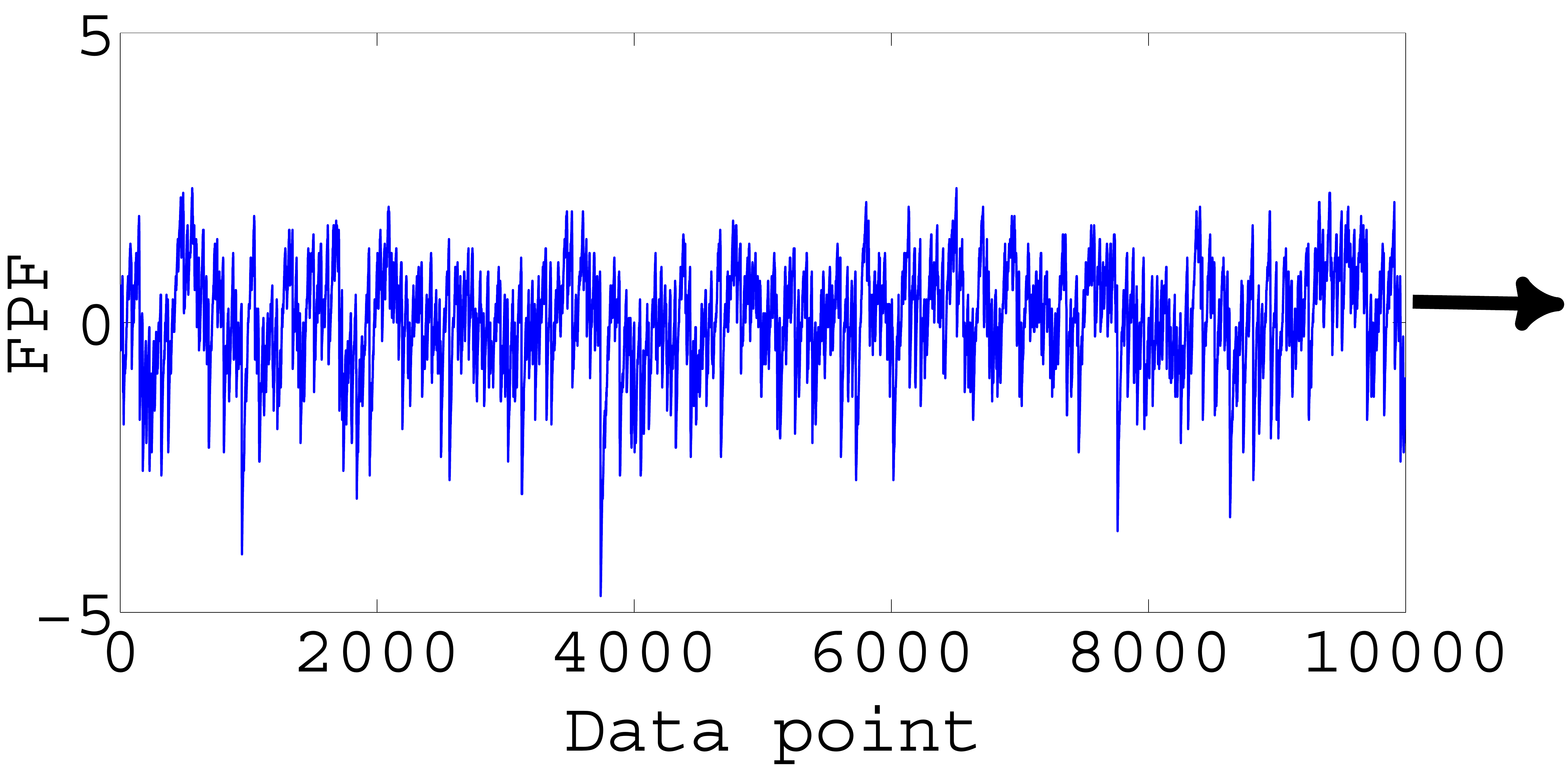}
\includegraphics[width=4.5cm, height=3cm]{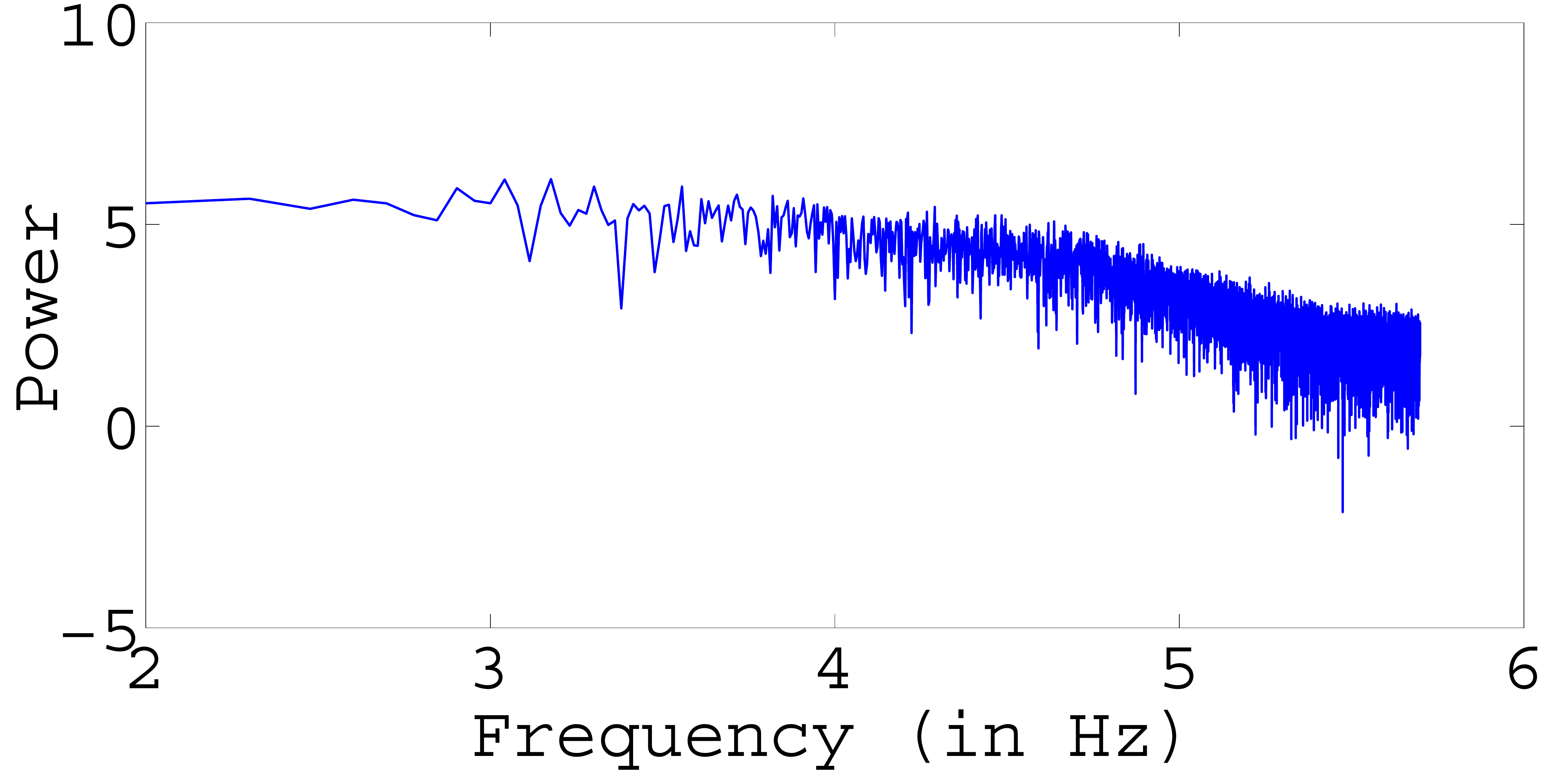}
\includegraphics[width=4.5cm, height=3cm]{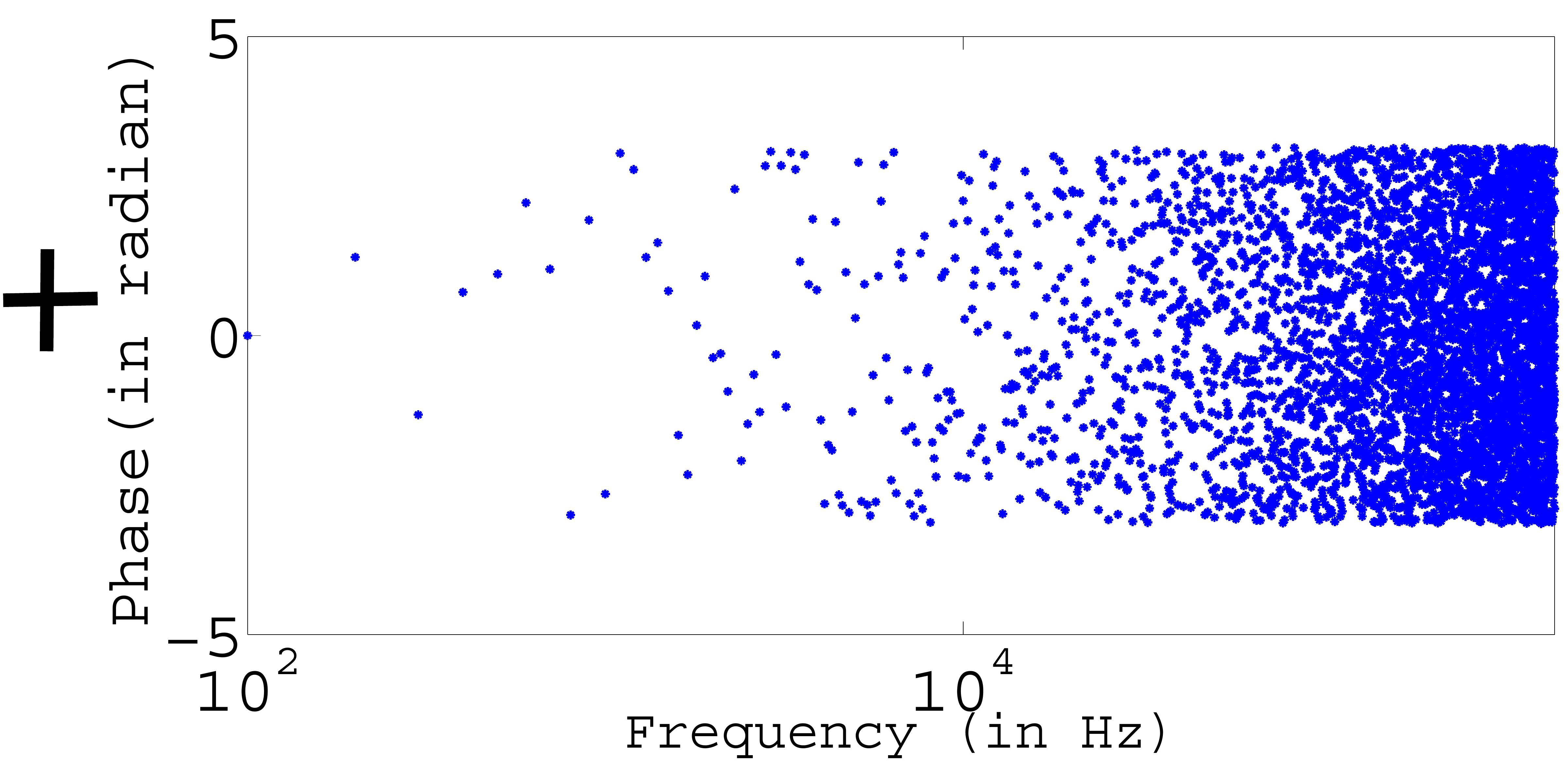}
\caption{Original raw signal decomposed into its amplitude and phase part }
\label{rap}
\end{figure}

When we attempt to obtain phase information from data, the Fourier transformation has been the starting point for this purpose. The Fourier transformation of a time series X(t)is defined as.

\begin{eqnarray}
X(\omega)=\int_{-\infty}^{\infty} X(t) e^{-i\omega t} dt
\label{1}
\end{eqnarray}

where $\omega$ is the angular frequency which provides information about the amplitude $X(\omega)$ and phase distribution
$\phi(\omega)=\arctan \frac {Im X(\omega)} {Real X (\omega)}$. The decomposition of a time series into its amplitude spectrum and phase part is shown in fig \ref{rap}. In space plasma research in the context of geomagnetic pulsation and the power law type spectrum of magnetic field turbulence \cite{rob} amplitude spectrum has always been in discussion in the literature for many years. The phase distribution on the other hand has not achieved much attention in space plasma application. A possible reason may be that the phase distribution in Fourier space appears to be completely random as seen from the phase part depicted in the right panel of Fig. \ref{rap}.


\begin{figure}
\centering
\includegraphics[width=8cm,height=4cm]{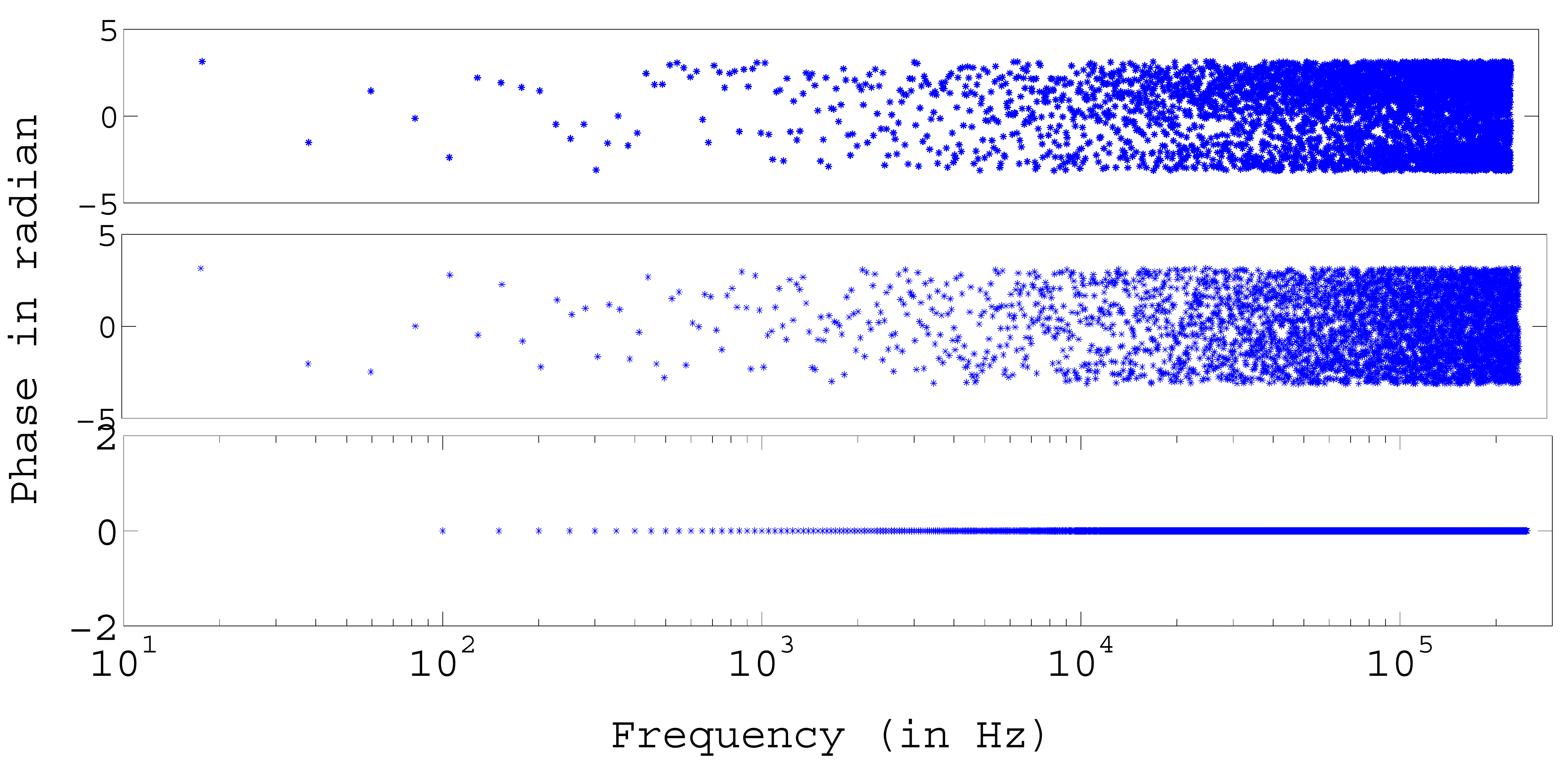}
\caption{Distribution of wave phases  for original FPF (ORG), phase-randomized surrogate(PRS) and phase-correlated surrogate (PCS) from upper to lower panel}
\label{phase}
\end{figure}

\begin{figure}
\centering
\includegraphics[width=8cm, height=5cm]{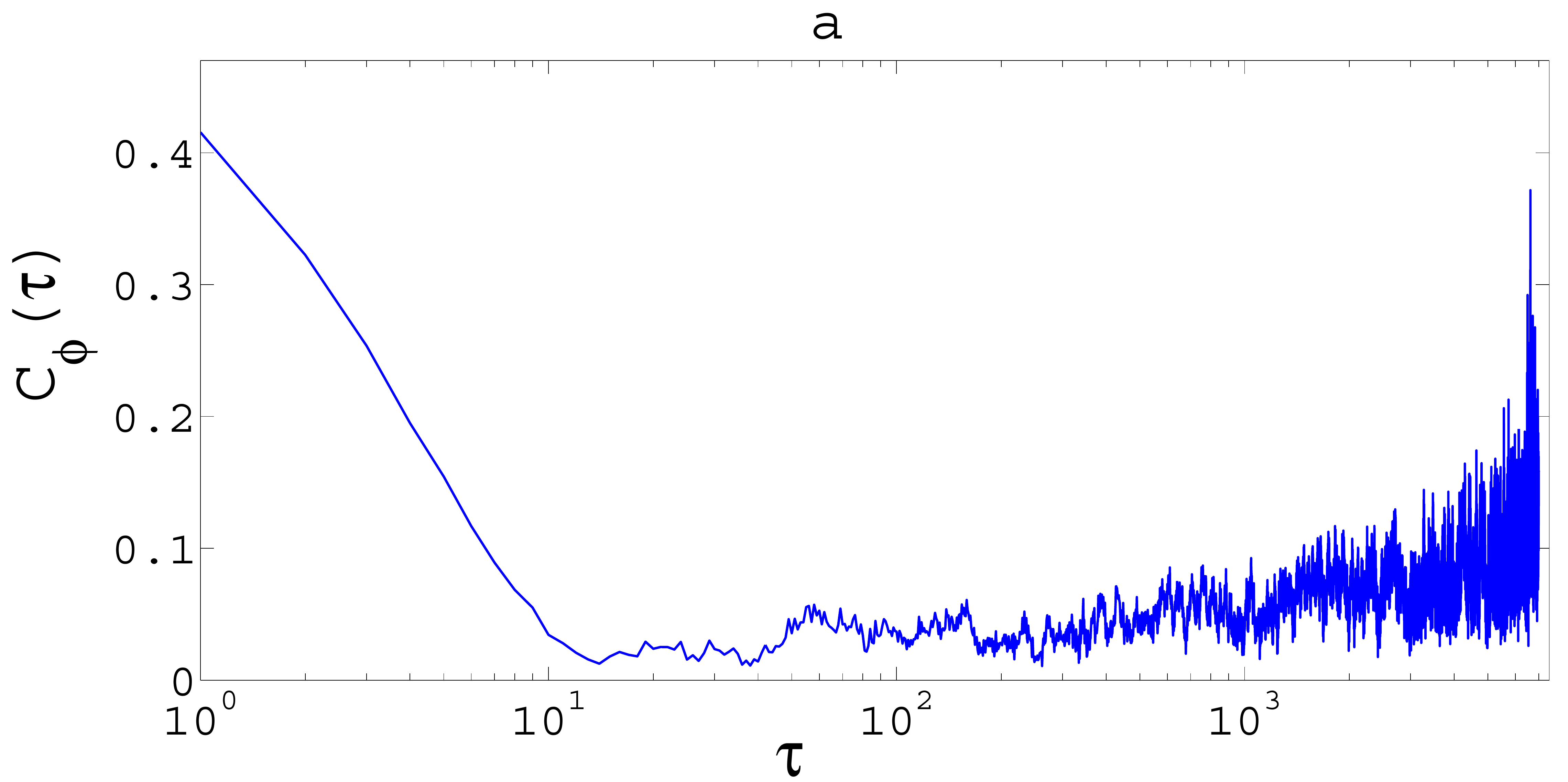}
\includegraphics[width=8cm, height=5cm]{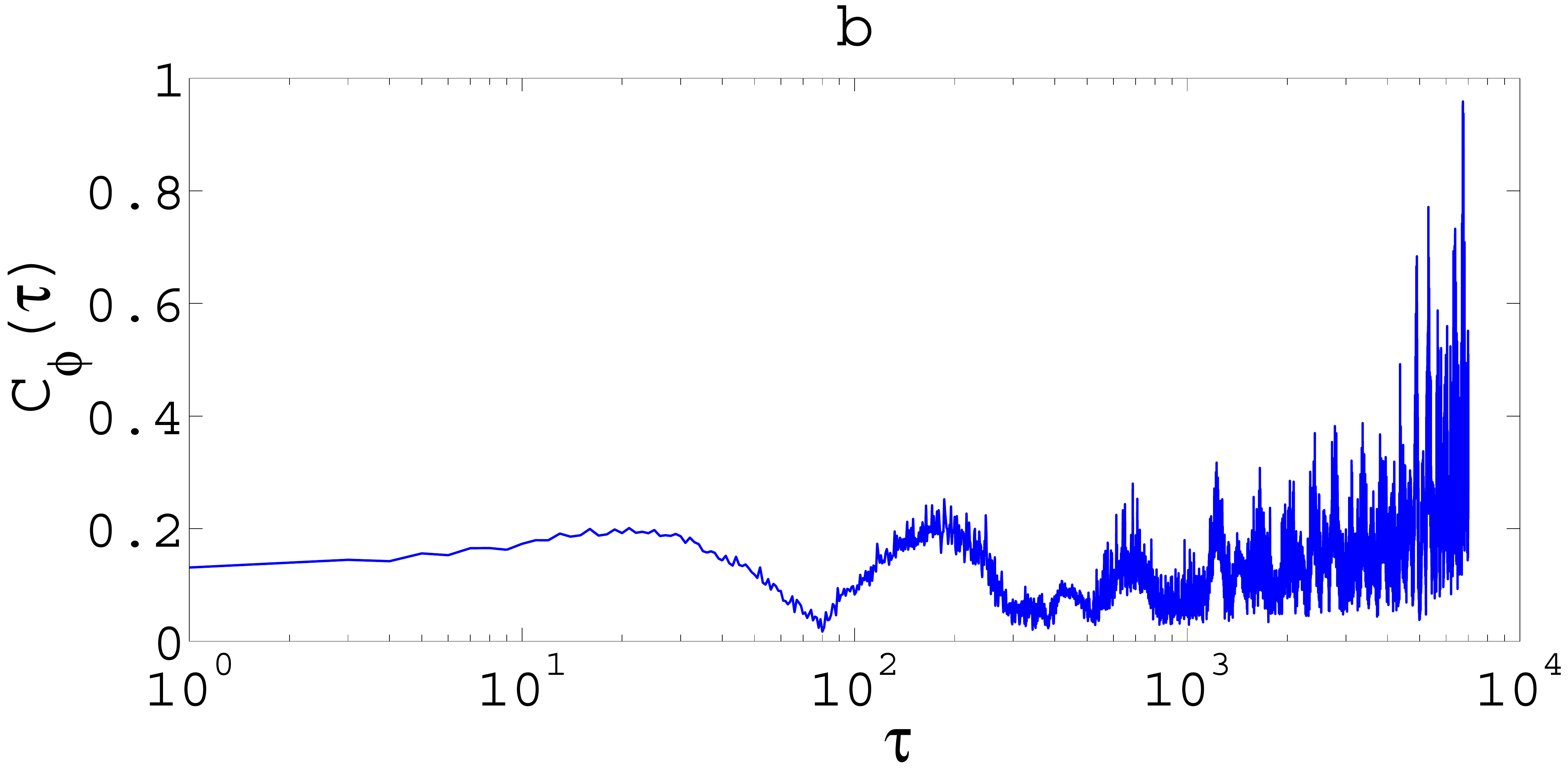}
\caption{Profile of phase coherence index with $\tau$ for a) $B_{T}=9.6G$  b) $B_{V+T}=12.63G$}
\label{ctau1}
\end{figure}

\begin{figure}
\centering
\includegraphics[width=14cm, height=8cm]{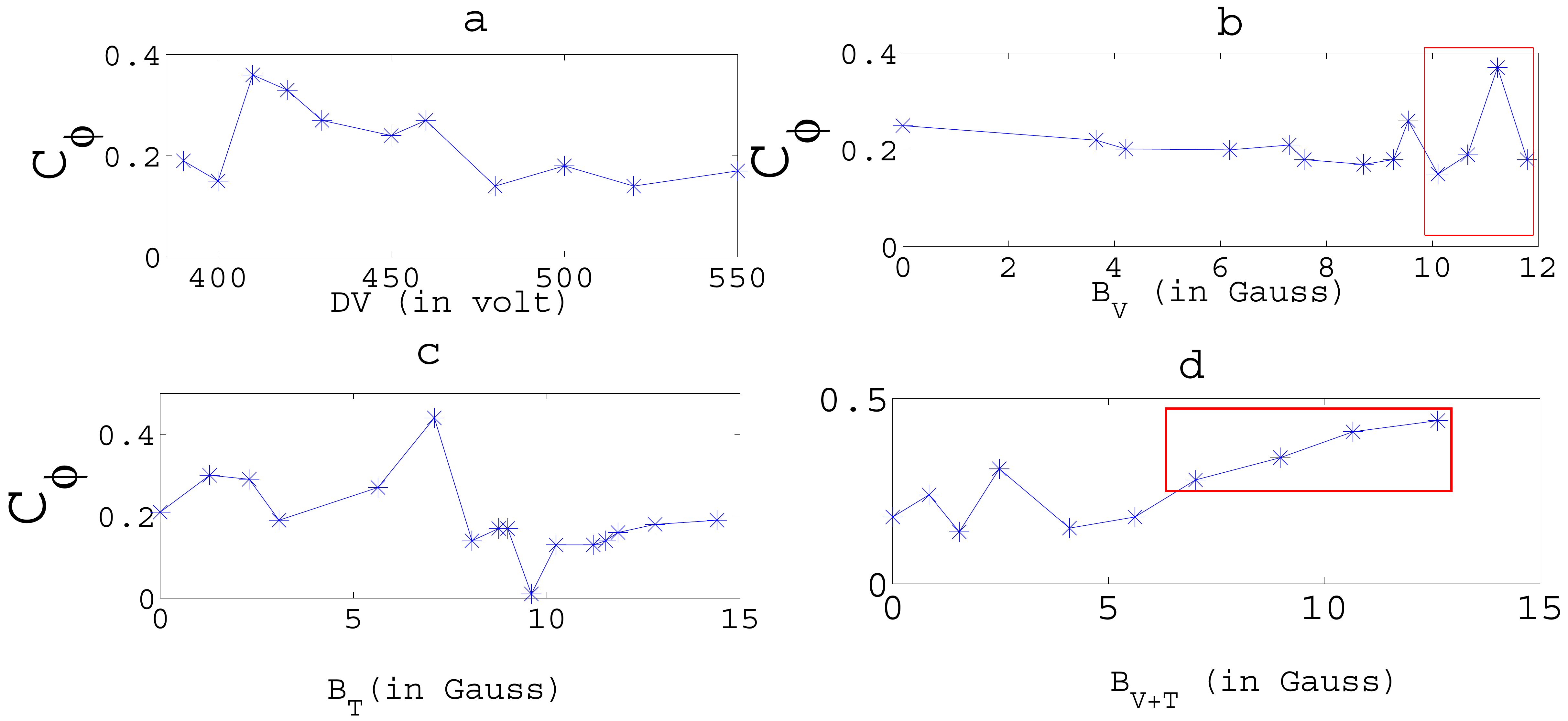}
\caption{Profile of phase coherence index with variation in a) Discharge voltage(DV) b) Vertical magnetic field ($B_V$) c) Toroidal magnetic field ($B_T$) d) Vertical field at fixed $B_T$ ($B_{V+T}$) }
\label{ctau}
\end{figure}

Koga et al. \cite{koga} introduced a method to evaluate the degree of phase coherence \cite{koga1} among the Fourier modes
quantitatively. From a given FPF we generate two sets of surrogate data. Firstly, we decompose the original data (ORG) into the power spectrum and the phases using the Fourier transform. We then randomly scramble the phases keeping the power spectrum unchanged, and from these two pieces of information in the Fourier space, we perform the inverse Fourier transform to create the phase-randomized surrogate (PRS). Phase-correlated surrogates(PCS)
are generated by making the phases equal without shuffling the Fourier phases. The three data, ORG, PRS and PCS share exactly the same power spectrum,
while their phase distributions are all different. The distribution of phases of the ORG, looks almost as random as that of the PRS in Fig. \ref{phase}, due to the arbitrary choice of the coordinate origin and so the phase coherence even if it exists is easily obscured. So, we can characterize the difference in the phase distribution by the difference in the waveform in real space, instead of dealing with the variables in Fourier space. The difference can be captured by the path length, the expression of which is given below in equation \ref{path}.

\begin{eqnarray}
S(L,m)=<|L(t_i+\tau)-L(\tau)|^m>
\label{path}
\end{eqnarray}

\begin{eqnarray}
C_{\phi}(\tau)=\frac{L_{PRS}(\tau)-L_{ORG}(\tau)}{L_{PRS}(\tau)-L_{PCS}(\tau)}
\label{path1}
\end{eqnarray}

where m is order of the path length and $\tau$ is the measure characterising the magnification level of the curve. We define the phase coherence index
to evaluate the degree of phase coherence ($C_{\phi}(\tau)$) in equation \ref{path1} between the Fourier modes. If the original data has random phase
then $C_{\phi}(\tau)$ would be 0 whereas $C_{\phi}(\tau)$=1 if the phases are completely correlated.
The profile of phase coherence index with variation in $\tau$ for the FPF's with $B_T$, $B_{V+T}$
is shown in Fig \ref{ctau1}. We can found that $C_{\phi}$ remains in the value from almost 0.01
to maximum 0.5 for almost wide range of the values of $\tau$ for $B_{V+T}$=12.63G whereas that takes the value from minimum 0.01 to 0.24 for $B_T$ of 9.6G as also ensured from the maximum value of $C_{\phi}(\tau)$ with $B_T$, $B_{V+T}$ as depicted in Fig. \ref{ctau}.

The plot of phase coherence with the variation in discharge voltage, vertical, toroidal, mixed magnetic field has been portrayed in Fig. \ref{ctau}. The value $C_{\phi}(\tau)$ takes it maximum when the phenomena of homoclinic transition occurs for increasing DV. For initial increase in vertical magnetic field the values of phase coherence index are shown to lie in constant range but the higher values of $B_V$ has led to the increase in $C_{\phi}(\tau)$ as marked in the figure where the shape of the glow starts becoming some sort of distorted elliptical from the original spherical shape. In the marked portion the maximum value correspond to the case where the toggling of the time scale in relaxation oscillations takes place i.e is the rise, decay time of the relaxation oscillation completely get changed as observed from Fig \ref{rawver}j-k. Gradual increase in value of $C_{\phi}(\tau)$ within the region from 6G to 11.22G shows similar trend with the dominant frequency unlike at $B_V$=11.78G where the value of phase coherence again get decreased. This smooth of increase in $C_{\phi}(\tau)$ correspond to the spreading of the anode glow to occupy both sides of the electrode (Fig \ref{glow1}).

The region of increasing phase coherence index with $B_T$ indicates the portion in Fig \ref{freqdom} where the dominant frequency started getting increased.
The emergence of a pear shaped glow different from the previous ones correspond to the maximum values of $C_{\phi}$. At a value of $B_T$=9.6G a sudden
appearance of very low amplitude FPF resulted in minimum correlation between the wave phase (lowest values of  $C_{\phi}(\tau)$) followed by the constant values of $C_{\phi}$ in high toroidal magnetic field region where the glow shapes are seen to remain same (Fig. \ref{glow2} f-h). Here also we can observe that the maximum phase coherence index is associated with the power/energy occupying in a larger region of frequency band. Finally the application of $B_V$
on fixed toroidal field i.e $(B_{V+T}$ lead to the smooth increase in the values of phase coherence index in a region marked by red rectangle implying the enhancement in the correlation between the wave phases with the slight increase in the intensity of the pear shaped glow. Gradually increasing $C_{\phi}$ values are seen to be prominent for the broadband fluctuations depicted in Fig \ref{rawmix} h-j. Thus the results prove that the values of phase coherence is indicative of the accumulation of power/energy in a band of frequency. So the effect of mixed field ($B_{V+T}$) is seen to be prominent in enhancing the phase coherence index as observed from Fig. \ref{ctau}.

\section{Detrended fluctuation analysis}

\begin{figure}
\centering
\includegraphics[width=9cm]{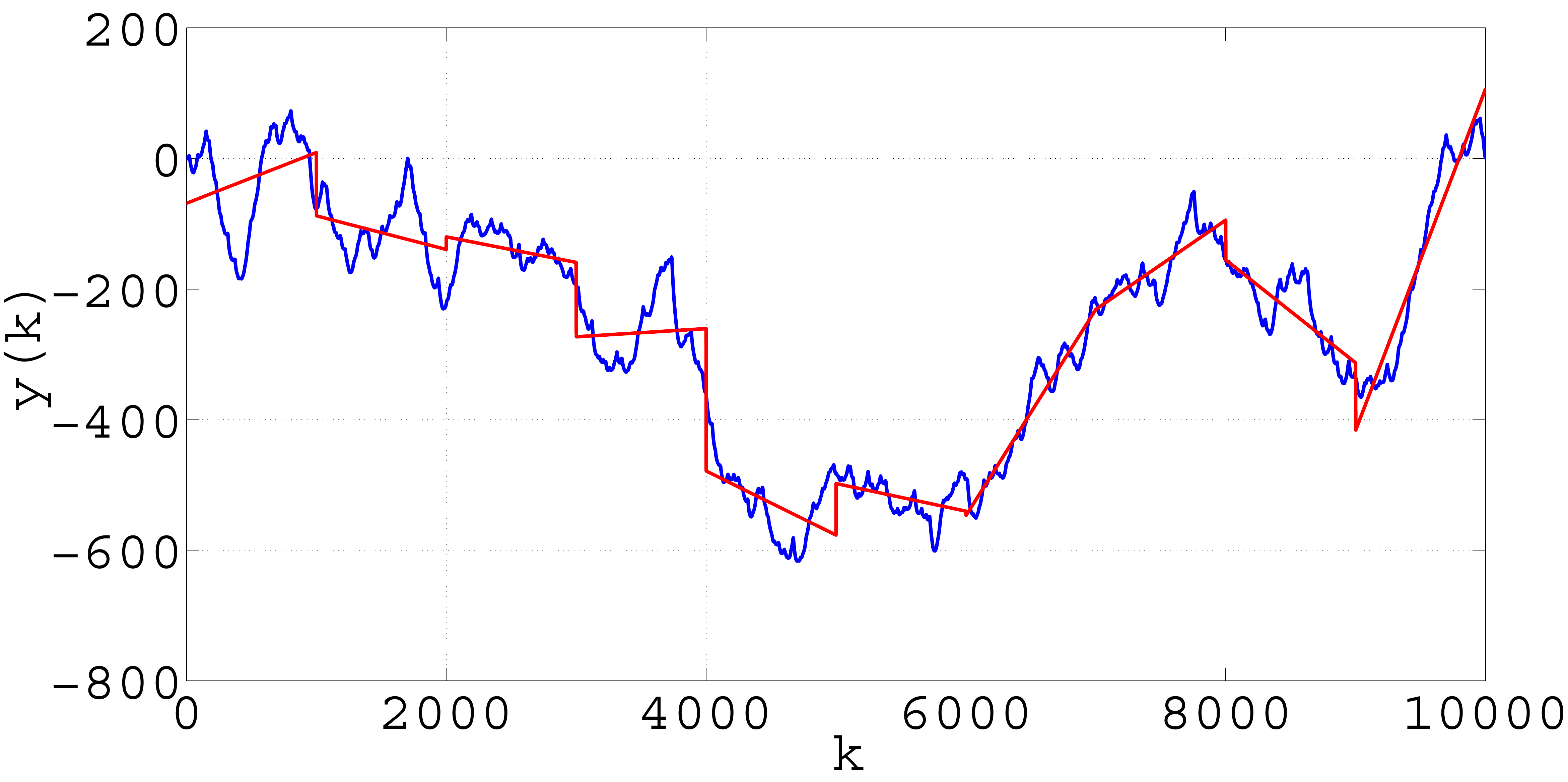}
\caption{Integrated time series superposed on least square fitted trend}
\label{DFA}
\end{figure}

To illustrate the DFA algorithm we use the time series shown in Fig \ref{DFA} (in left panel) as an example along with the integrated time series with
the solid red lines indicating the trend estimated in each box by a least square fit. Following the approach adapted by Peng et al. \cite{peng}
we first integrate the time series $ y(k)=\sum_{i=1}^k [x(i)-x_{mean}]$ followed by the dividing of the time
series into boxes of equal length n. A least square line representing the trend is fitted to the data.
A y coordinate of the straight line segment is denoted by $y_n(k)$ in each box. Next we detrend the integrated time series y(k)
by subtracting the local trend, $y_n(k)$ in each box. Root mean square fluctuation of this integrated and detrended time series is calculated by

\begin{eqnarray}
F(n)=\sqrt{\frac{1}{N}{\sum_{k=1}^N[y(k)-y_n(k)]^2}}
\label{diagline}
\end{eqnarray}

The process is repeated over all time scales to provide a relationship between the average fluctuation as a function of box size F(n) and the box size n.
A linear relationship on a log log graph indicates the presence of scaling with scaling exponent being $\alpha$ implying that $F(n) \sim n^{\alpha}$. Value of
$\alpha$ greater than 0.5 and less than or equal to 1.0 indicates persistent long range correlation and that lying in the range $0<\alpha<0.5$ represent
different type of power law correlation such that large and small values of the time series are likely to alternate.

\begin{figure}
\centering
\includegraphics[width=8cm, height=5.5cm]{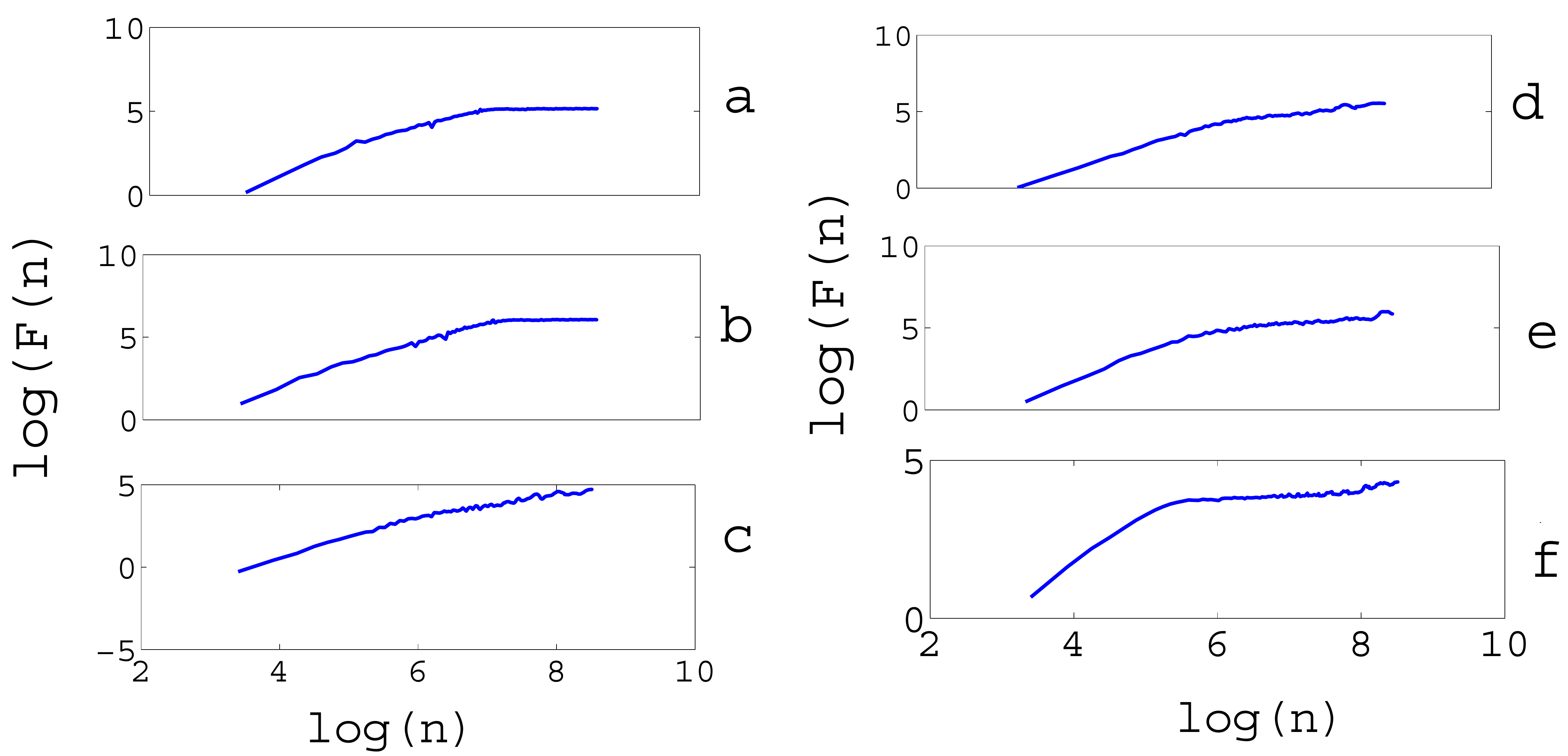}
\includegraphics[width=8cm,height=5.5cm]{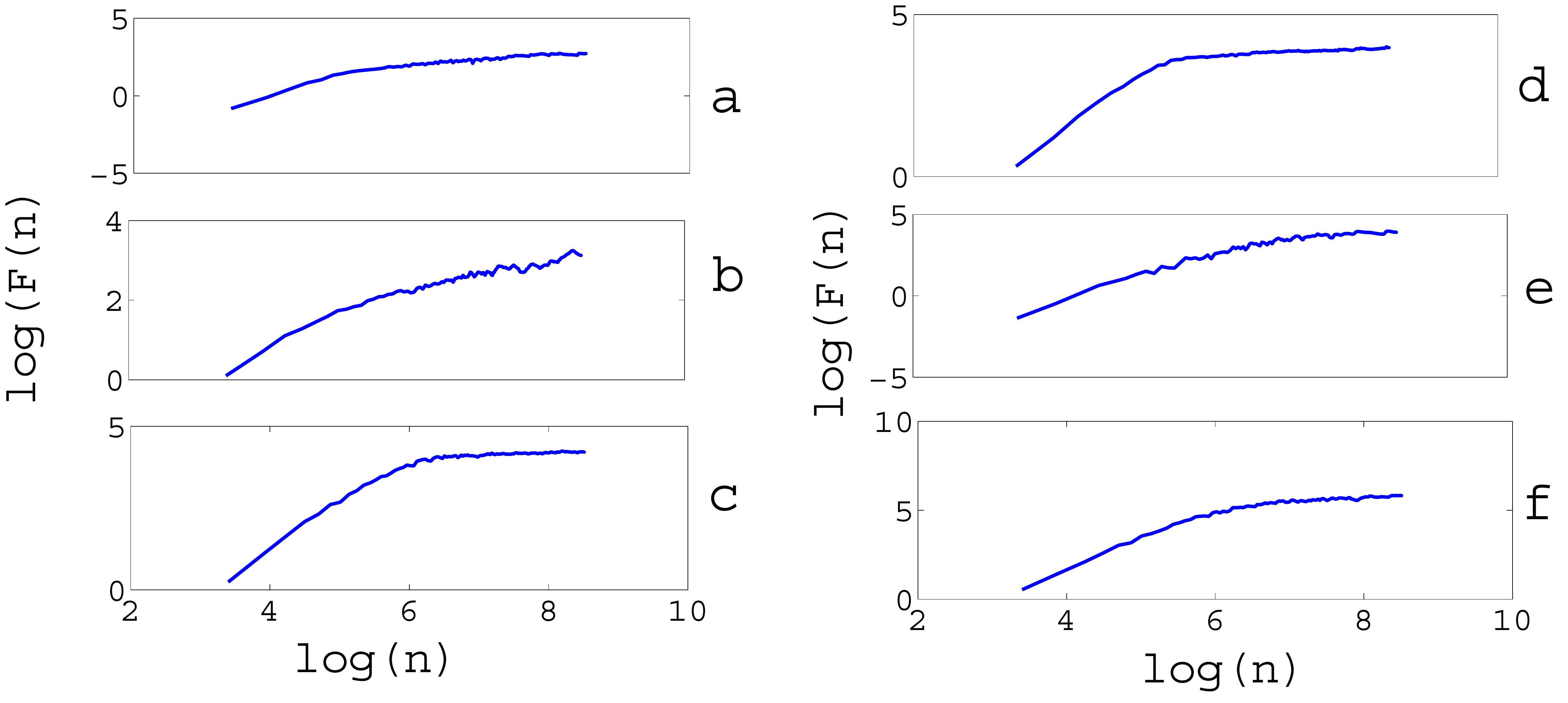}
\caption{Plot of $logF(n)$ vs $log(n)$ for increasing toroidal voltage a) 0G b) 1.28G c) 5.63G d) 7.09G e) 8.06G f) 8.98G in the left panel and a) 9.34G b) 10.24G c) 11.52G d)11.84G e) 12.8G f) 14.4G}
\label{DFA1}
\end{figure}

\begin{figure}
\centering
\includegraphics[width=9cm, height=6cm]{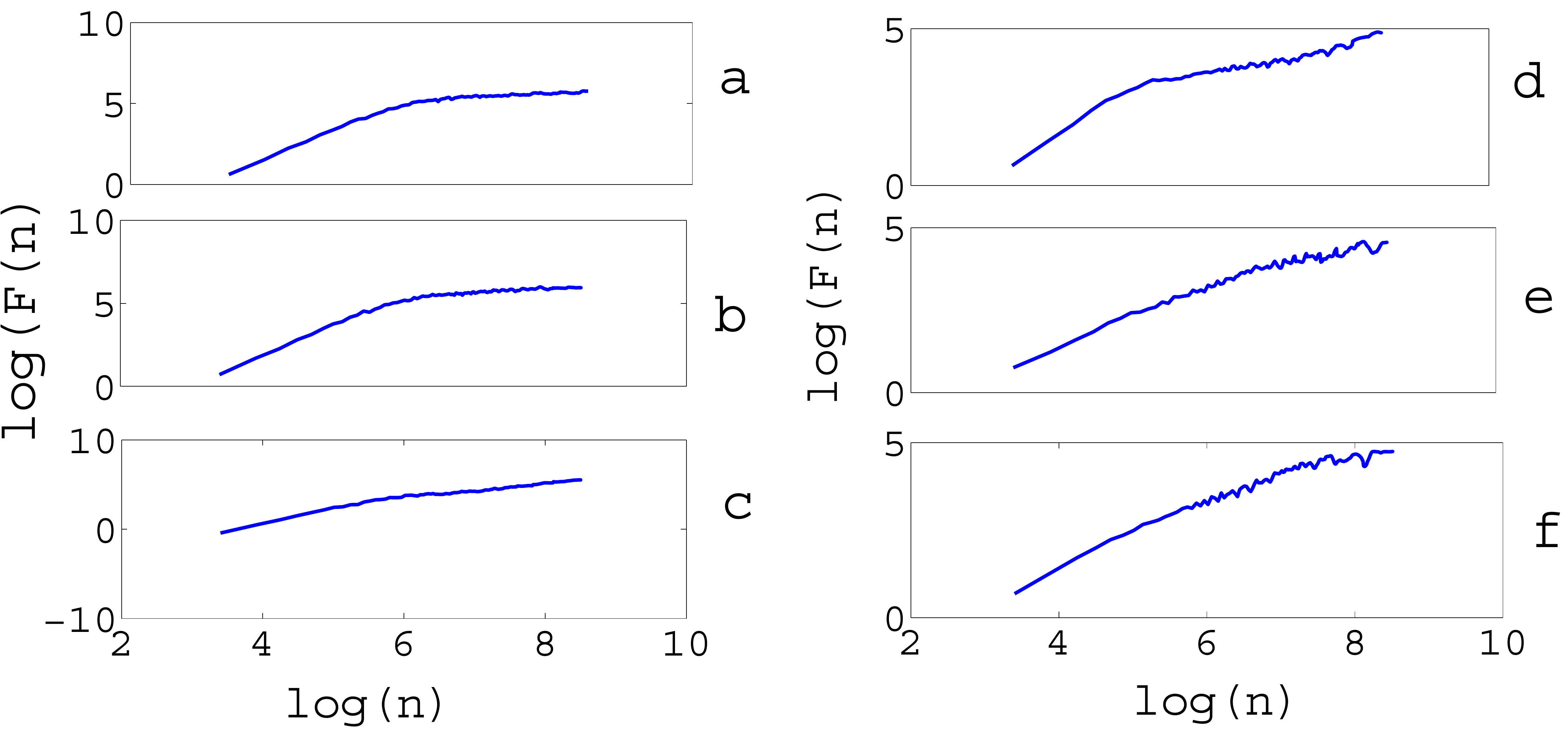}
\caption{plot of $logF(n)$ vs $log(n)$ increasing $ B_V$ at fixed $B_T$ ($B_{V+T}$) a) 0.84G b) 2.47G c) 5.61G d) 7.01G e) 10.66G f) 12.63G }
\label{DFA2}
\end{figure}

We illustrate results using the above mentioned technique on the fluctuations acquired with increasing $B_T$, $B_V$, $B_{V+T}$ respectively in Fig. \ref{DFA1}, Fig. \ref{DFA2}. The slope of the curves allows us to check for the scaling exponent
after carrying out DFA analysis. DFA performed on the fluctuations in the left panel of fig \ref{DFA1}
with toroidal field upto 8.98G yield the values of $\alpha$ to be 1.28, 1.31, 0.99, 1.24, 1.27, 1.25. Power law exponent values greater than 1 indicate
the existence of perfect correlated dynamics. Altogether the exponent is viewed as the roughness of the time series, the larger the value of coefficient
$\alpha$ the smoother will be the time series. The hint of long range correlated dynamics ($0.5<\alpha<1$) is found for $B_T$=5.63G where the corresponding FPF abruptly change its nature of relaxation oscillation. The prominent signature of the existence of long range behaviour has been observed from the slope of the curves in right panel of Fig \ref{DFA1}. The values of the estimated scaling exponent are found to lie within $0.5<\alpha<1$ for DFA executed on FPF's
acquired under high value of $B_T$ from 9.34G to 14.4G. The application of $B_{V+T}$ yield the values of scaling exponents of 1.36, 1.23, 0.91;(1.09,0.54); 0.72,0.78, 0.75 indicating a toggle from perfect correlated dynamics to a long range persistence. A phenomena worth observing is the prominent region of double scaling ($\alpha$=1.09, 0.54) which is observed for $B_V$=11.2G followed by the long range behaviour. So the appearance of long range behaviour is noted only under the application of high $B_T$ or the mixed field $B_{V+T}$. The values of the scaling exponent can also be associated with the anode glow and phase coherence index with the persistence occurring during the appearance of completely new inverted pear shaped glow after the phenomena of self rotation of the glows with $C_\phi$ being maximized during the long rang persistence. Similarly the scaling exponent values indicating long range persistence at higher values of $B_{V+T}$ correspond to the gradually increasing trend of $C_\phi$. Now the estimated values of the scaling exponent under the effect of increasing DV and $B_V$ are listed in table \ref{tab}. All the values of the scaling exponent are seen to lie above 1 indicating perfect correlated dynamics and absence of any long range persistence.



\begin {table}
\caption {Results of the quantitative values of the scaling exponent for increasing discharge voltage (DV) and vertical magnetic field ($B_V$) } \label{tab}
\begin{center}
\begin{tabular}{| l| l| l|l|}
\hline
  DV (in V)  &  \ \ \ \       slope($\alpha$)  & \ \ \ \      $B_V$ (in G)   & \ \ \ \  slope($\alpha$)     \\  \hline
  390        &  \ \ \ \            1.22       & \ \ \ \        3.64         & \ \ \ \   1.37   \\  \hline
  410        &  \ \ \ \            1.14       & \ \ \ \        6.17         & \ \ \ \   1.39   \\  \hline
  430        &  \ \ \ \            1.40       & \ \ \ \        8.70         &  \ \ \ \  1.36   \\  \hline
  450        &  \ \ \ \            1.78       & \ \ \ \        9.54         & \ \ \ \   1.70   \\  \hline
  460        &  \ \ \ \            1.79       & \ \ \ \        10.66        & \ \ \ \   1.53   \\  \hline
  480        &  \ \ \ \            1.81       & \ \ \ \        11.78        & \ \ \ \   1.54   \\  \hline
 \end{tabular}
\end{center}
\end {table}

\section{conclusion}

 To summarize, the study of the interplay of oscillations with parameters like DV, $B_V, B_T, B_{V+T}$  has been carried out with the simultaneous observation of the dynamics of the fireball. Estimation of the phase coherence index to characterise the correlation between the modes has been accomplished. The scaling exponent associated with this FPF's has been illustrated using the DFA algorithm. Interplay of transitions between oscillations lead to the toggling from correlated dynamics to long range persistence as ensured from the results of DFA. The change in the dynamics of the fireball along with the associated phase coherence index values has been studied in detail. The results of finite $C_{\phi}$ demonstrate the existence of finite phase correlation indicating the nonlinear wave interaction in process. For the case of increasing DV we found the maximum correlation at the instant of the occurrence of homoclinic transition. Display of dominant frequencies with the variation in $B_V, B_T, B_{V+T}$ has been carried out to understand about the frequency information. A wide range of dominant frequencies are noted for increasing toroidal and mixed magnetic fields with the latter following the increasing trend. Existence of power/energy concentration in a large region of frequency band is thought to be attributed to the gradual increase in phase coherence index values for those fields whereas for $B_V$ the $C_{\phi}$ values are found to associated with the change in the time scale of relaxation oscillation. Talking about scaling exponents, the evidence of long range correlated dynamics is noted for higher values of $B_T$ whereas the toggling from perfect correlated dynamics to long range persistence via double scaling region has been observed for $B_{V+T}$.

 Recalling the fireball phenomena electrons accelerated in the sheath of the positively biased electrode excite or ionize the neutrals leading to
 the expansion of the sheath into a double layer of potential just above the ionization potential. The common assumption that a fireball can collect as many electrons emitted from the cathode is not correct since secondary electrons in the discharge can also be collected and balanced by ions flowing to the chamber walls. In our experiment, emergence of inverted pear shaped glow upon the application of $B_T$ correspond to the maximum correlation between phases quantified by maximum $C_{\phi}$ and long term persistence. The smooth increase in $C_{\phi}$ for higher $B_V$ occurs during the spreading of glow on both sides of the electrode. When the plasma production and losses get out of the balance the fireballs can undergo relaxation oscillation. Growth and collapse of the fireball can be considered runaway processes whose time scales are governed by the ion transit time through the fireball. The recovery process depends on the density replenishment from the background plasma which may take longer due to the lower density and larger scale. In the context of the shape of the fireball we know that the fireball structure must be in a force balance $m_e n_e v_e$=$m_i n_i v_i$ otherwise it would not be a stationary structure. In a uniform unmagnetized plasma with all radial forces cancelling each other fireballs of spherical or cylindrical geometries can be obtained.

Scaling exponent using DFA helps us to find the long range persistence, correlated dynamics. The persistence long range behaviour in presence of higher $B_T$ corresponds to the pear shaped glow depicted in Fig. \ref{glow2}e-h. The shape of this particular nature of fireballs also indicates the presence of long range correlated dynamics in case of increasing $B_{V+T}$. So understanding the dynamics of the oscillation through the techniques like DFA, $C_{\phi}$ is a crucial step for the characterisation of the device for application purposes along with the dynamics of the fireball.

\noindent {\textbf{Acknowledgments:}} The author would like to acknowledge the Director of SINP for his constant support and two project students Aishik Biswas and Dipankar Ganguly from techno india college, saltlake for their assistance during the experiment.

\end{document}